\documentclass[]{IEEEtran}
\usepackage{cite}
\usepackage{amsmath,amssymb,amsfonts,mathtools}
\usepackage{bbm, bm}
\usepackage{algorithm}
\usepackage{algcompatible}
\usepackage{graphicx}
\usepackage{textcomp}
\usepackage{xcolor}
\usepackage{subcaption}
\usepackage{siunitx}
\usepackage{hyperref}
\usepackage{flushend}

\newcommand{\grad}{\nabla}

\newcommand{\bI}{\mathbf{I}}

\newcommand{\bzero}{\mathbf{0}}

\newcommand{\bx}{\mathbf{x}}

\newcommand{\bz}{\mathbf{z}}

\newcommand{\bepsilon}{{\boldsymbol{\epsilon}}}

\def\BibTeX{{\rm B\kern-.05em{S  i\kern-.025em b}\kern-.08em
    T\kern-.1667em\lower.7ex\hbox{E}\kern-.125emX}}
    
\definecolor{mehdi}{RGB}{0,0,250}

\definecolor{Samad}{RGB}{0,250,0}

\definecolor{Matti}{RGB}{250,0,0}

\graphicspath{ {./Figures/} }
    
\begin{document}

\title{Generative AI-Based Probabilistic Constellation Shaping With  Diffusion  Models
}

\markboth{}
{}
\author{\IEEEauthorblockN
{Mehdi Letafati, \IEEEmembership{Student Member, IEEE,}
			Samad Ali, \IEEEmembership{Member, IEEE,}   and
		Matti Latva-aho,
		   	\IEEEmembership{Senior Member, IEEE}
}
\textsuperscript{}\thanks{ 
This work has been submitted to the IEEE for possible publication.  Copyright may be transferred without notice, after which this version may no longer be accessible. 

Preliminary results of this paper are submitted to the IEEE International Conference on Machine Learning for Communication and Networking  (ICMLCN 2024), Stockholm, Sweden, May 2024 \cite{conf_constell}.  

The authors are with the Centre for Wireless Communications, University of Oulu,
Oulu, Finland (e-mails: mehdi.letafati@oulu.fi; 
 samad.ali@oulu.fi; matti.latva-aho@oulu.fi). 
}}

\maketitle

\begin{abstract}   
Diffusion models are at the vanguard of  generative AI  research 
with renowned  solutions  such as  ImageGen  by  Google Brain and   DALL.E 3  by OpenAI.  
Nevertheless,  the potential merits of diffusion models for communication  engineering applications are not fully understood yet. 
In this paper, we aim to  unleash the  power of generative AI  
for PHY design of constellation symbols  in communication systems.   
Although  the geometry of constellations is  predetermined  according to  networking  standards,   e.g., quadrature amplitude modulation (QAM),  probabilistic 
shaping can design the probability of occurrence (generation) of constellation  symbols. This can help improve  the information rate and decoding performance of communication systems.  
We exploit   
the ``denoise-and-generate'' characteristics  of 
denoising diffusion probabilistic models (DDPM) for probabilistic constellation shaping.   The key 
idea is to learn generating  constellation symbols out of
noise, ``mimicking'' the way the receiver performs symbol reconstruction.  
This way, we make the constellation symbols sent
by the transmitter, and what is inferred (reconstructed) at the
receiver become as similar as possible, resulting in as few mismatches  
as possible.  
Our  results 
show  that the generative AI-based  scheme  outperforms 
deep neural network (DNN)-based 
benchmark 
and uniform shaping, while providing \emph{network resilience} as well as \emph{robust out-of-distribution
performance} under low-SNR regimes and  non-Gaussian assumptions. Numerical  evaluations    highlight  $30\%$ improvement in terms of cosine similarity and a threefold improvement in terms of mutual information compared to  DNN-based approach for 64-QAM geometry.  
\end{abstract}


\section{Introduction}\label{sec:Intro}  
With the incredible results achieved from generative 
pre-trained transformers (GPT) and diffusion models, 
generative AI  is envisioned to be at the forefront of technological advancements   in various industrial and academic domains \cite{Petar, NVIDIA, WiGenAI}.    In the vanguard of generative AI  research,  diffusion models 
have 
showcased  outstanding  performance    
with renowned   solutions such as  
ImageGen,\footnote{https://imagen.research.google/}  
  DALL.E 3,\footnote{https://openai.com/dall-e-3} 
 and stable diffusions,\footnote{https://github.com/CompVis/latent-diffusion}
   to name a few \cite{DM_survey}.    
In parallel, the future generations of wireless systems  entail the extensive integration of AI and machine learning  (AI/ML) algorithms into the  communication and networking design, realizing   ``AI-native'' wireless  networks \cite{3gpp, hexaX, Samad, Nokia_white, twelve_6G}. 
This highlights the necessity for developing novel  AI/ML solutions that are  tailored to address the requirements of emergent  communication scenarios.  

The majority of the research carried out  thus far on AI-native  communication and signal  design,  has primarily concentrated on ``discriminative'' AI/ML models    
\cite{DNN, DeepRx,  DeepTx, geom_arXiv}.  
The  objective  of such models is to simply  learn the ``boundaries’’ between classes or latent spaces within  high-dimensional signals. In contrast,  ``generative'' models essentially  learn the  \emph{representations}  
of highly-structured signals and generate  desired samples accordingly.  
{Inspired by the insights provided in \cite{WiGenAI} for WiGenAI (a new vision on generative AI-based wireless system design),  our goal in this paper is   
to unleash the power of generative AI  at the transmitter of wireless systems,  by  proposing  diffusion-based  probabilistic  constellation shaping.} 
\emph{To the best of our knowledge, this is the first technical paper that proposes diffusion models for   the application  of   
PHY signal design in wireless communications.}

\subsection{Literature Review}
Despite the fact that  remarkable results have been achieved by diffusion models in  various  domains of  computer science,  such as natural language processing (NLP) \cite{DM_NLP}, computer vision \cite{DM_CV}, molecule generation \cite{DM_GNN},  and medical imaging \cite{DM_MRI}, 
 {there are only a few papers in communication literature that have studied the applications of diffusion models for wireless systems} \cite{CDDM, DDPM_conf_hwi, hybrid, hybrid2, CGM_ChanEst_WCNC}.      

In \cite{hybrid}, diffusion models are employed at the receiver of a communication system for image transmission,  as a complement for  deep learning-based joint source-channel
coding scheme.  
The goal is to progressively refine the image at the receiver, taking the perception-distortion trade-of into account.  
Considering a similar  objective within  a similar application, 
the authors in \cite{hybrid2}  combine  diffusion models  and invertible neural networks for high-quality source image recovery. 
The results highlight a perceptual  improvement in the  reconstruction  performance under low bandwidth and low signal-to-noise ratio (SNR) conditions, compared to the  deep learning-based approaches. 
As another hybrid approach,  \cite{CDDM} employs  diffusion models to improve the  receiver's performance  in terms of  channel estimation error removal.  
Although diffusion models have shown promising performance in removing noise components and  reconstructing data samples,  
the authors employ an  additional autoencoder  block.  However, implementing  two different ML models, each with a distinct objective function  
can impose computational overhead  to the network.  
Furthermore,  the output signals of the employed encoder part does not necessarily follow the standard format of constellation symbols, making the scheme inapplicable to real-world wireless systems.     
Score-based diffusion models  are employed in \cite{CGM_ChanEst_WCNC} for  channel estimation in  multi-input-multi-output (MIMO) wireless communications. RefineNet neural architecture \cite{RefineNet}, comprised of 24 hidden channels in the first layer and a depth of six residual blocks in both the encoder and decoder parts, is implemented to estimate the gradient of the log-prior of wireless channels. 
The results  imply  a competitive performance of diffusion models  for both in-distribution  and out-of-distribution (OOD) scenarios  
compared to generative adversarial network (GAN).  
In a recent work \cite{DDPM_conf_hwi}, we have  proposed exploiting diffusion models  for a hardware-impaired communication system, showcasing  more than $25$ dB
improvement in reconstruction performance compared to deep neural network (DNN)-based receivers.

\subsection{Our Work}  
With the aid of generative AI, particularly diffusion models,  our goal in this paper  is to take a step towards  a  {generative AI-native} system, in which  we can continuously design  radio signals, adapt to changes, 
and  realize ``reciprocal  understanding'' between  communication parties.   
To this end, we  propose denoising diffusion probabilistic models (DDPM),  as one of the state-of-the-art generative models  proposed by  Ho \emph{et al.} in 2020 \cite{DM_Ho},  for probabilistic  constellation shaping at the transmitter.   
To the best of our knowledge, {this is the first paper that proposes diffusion models for  
constellation shaping in communication systems.}     
 
The motivation behind our work is  
that the choice of constellations can significantly affect  the performance of communication systems.   
Recently, deep learning  techniques are proposed  for geometric shaping \cite{constell, opt_pulse_shape,waveform_learn}. They typically employ ``discriminative'' models, particularly  classical  autoencoders,   
and let the neural model decide about the constellation symbols for transmission.  This results in  arbitrary forms of constellation points that might not be compliant with wireless standards such as the 3rd Generation Partnership Project (3GPP) \cite{3gpp_constell}.         
To overcome this challenge,  probabilistic constellation shaping can be employed---it designs  the probability of occurrence (generation) of  constellation  symbols within the corresponding geometry, aiming to  enhance the information rate and decoding performance of communication systems.   
Unlike previous works that try to deal with  the optimization of discrete distributions \cite{constell}, we offer  a radically different approach and exploit   
the ``denoise-and-generate'' characteristic  of DDPMs for probabilistic   shaping.

In our proposed scheme,  a DDPM   
is first trained with the goal of learning the diffusion process for generating constellation symbols.  
Within each transmission slot,  the transmitter 
runs the diffusion model to probabilistically shape (generate) the constellation symbols  according to the signal-to-noise ratio (SNR) level.  
Intuitively, the goal is to do shaping in a way that the information-bearing constellation symbols generated at the transmitter, and what is inferred (reconstructed) at the receiver become as similar as possible,  resulting in as few mismatches between the communication parties as possible.  
The key idea to fulfill this requirement is that   
we  exploit the ``denoise-and-generate'' characteristic of DDPMs
to help communication parties maintain 
``reciprocal understanding'' of how to map and demap the information symbols according to the noise level (or equivalently the channel SNR) of the communication link over time.   
More details of our proposed approach  are provided in Section \ref{sec:SysMod}. 
{
Previous works  
require 
both the transmitter and the receiver to get involved in the ``joint’’ training of the system    
by  
passing the loss function gradient  \cite{waveform_learn , constell } 
(or its approximation \cite{DNN}) through the channel layers. This might cause some incompatibility   issues between the transmitter (e.g., the base station of a cellular network from a specific vendor) and the receiver  (e.g., a user equipment from another vendor).   
Nevertheless, our approach requires only  one neural model which  is pre-trained and identically employed by both of the  communication parties within a diffusion-based framework.   
}
 
Through extensive  numerical experiments, 
we show that our proposed approach  outperforms DNN-based scheme with trainable constellation layer and neural demapper  \cite{constell}. Notably, we show that  $30\%$ improvement in terms of cosine similarity and a threefold improvement in terms of mutual information are  achieved compared to  DNN-based solution for $64$-QAM geometry and $0$ dB SNR. 
Our results also highlight that the proposed DDPM-based  scheme is  \emph{resilient} against  low-SNR regimes. 
We also study  the \emph{out-of-distribution (OOD) performance} of our scheme  under non-Gaussian assumptions.      
Furthermore, we compare our probabilistic shaping scheme with  uniform shaping baseline,  and show that the proposed DDPM achieves  a notable  performance compared to the conventional uniform shaping as well.

\subsection{{Paper Organization and Notations}}
In what follows, we first introduce the framework of DDPM,  together with the main formulas and the corresponding loss functions in Section \ref{sec:DDPM}. System model and our  proposed scheme are addressed in Section \ref{sec:SysMod}.  Furthermore,  the neural network  architecture and our  proposed DDPM-based algorithms  for probabilistic constellation shaping  are addressed in this section.  
Numerical   results are studied in Section \ref{sec:Eval}. Finally, Section \ref{sec:concl} concludes the paper.  
\subsubsection*{Notations} 
Vectors and matrices are represented, respectively,  by bold lower-case and upper-case symbols. $|\cdot|$ and $||\cdot ||$ respectively denote the  absolute value of a scalar variable and the $\ell_2$ norm of a vector. Notation $\mathcal{N}(\mathbf{x}; \boldsymbol{\mu}, \mathbf{\Sigma})$  stands for the multivariate normal distribution  with mean vector $\boldsymbol{\mu}$ and covariance matrix $\mathbf{\Sigma}$ for a random vector $\mathbf{x}$. Similarly, complex normal distribution with the corresponding mean vector  and covariance matrix is denoted by $\mathcal{CN}(\boldsymbol{\mu}, \mathbf{\Sigma})$. Moreover, the expected value of a random variable (RV) is denoted by $\mathbb{E}\left[\cdot\right]$   Sets are denoted by calligraphic symbols.  $\bm 0$ and $\bf I$ respectively show all-zero vector and identity matrix of the corresponding size. Moreover, $[N]$,  (with $N$ as integer) denotes the set of all integer values from $1$ to $N$, and $\mathsf{Unif}[N]$ (for $N > 1$) denotes discrete uniform distribution  with samples between $1$ to $N$. Also,  $\delta(\cdot)$ denotes the Dirac function.

\section{Preliminaries}\label{sec:DDPM} 
Diffusion models break down the data generation process into a series of incremental  ``denoising'' steps, during 
which the model corrects  itself, until ultimately  generating the desired samples.   
A diffusion model encompasses  two  processes, i.e.,   the forward and the reverse  diffusion processes.  
In the forward diffusion process, Gaussian kernels are applied to the data samples until they follow an isotropic  Gaussian  distribution, while  in the reverse process, the objective is to ``denoise'' and generate  desired  samples out of noise.

Consider ${\bf x}_0$ to denote  a data sample from the ground-truth probability distribution ${ q}({\bf x}_0)$. 
The forward diffusion process can be modeled by a conditional probability distribution $q({\bf x}_t|{\bf x}_{t-1})$ which applies the Gaussian kernel to data samples at each time-step 
$t \in [T]$,  with ${\bf x}_t$ denoting the diffused data sample at the $t$-th time-step, and $T$ denoting  the total  number of steps during which the diffusion process is carried out.      
Mathematically speaking, the forward diffusion process is expressed  as 
\begin{align}
    q(\mathbf{x}_t \vert \mathbf{x}_{t-1}) 
    & \sim \mathcal{N}(\mathbf{x}_t; \sqrt{1 - \beta_t} \mathbf{x}_{t-1}, \beta_t\mathbf{I}),  \label{eq:fwd_diffusion}  \\
q(\mathbf{x}_{1:T} \vert \mathbf{x}_0) 
& = \prod^T_{t=1} q(\mathbf{x}_t \vert \mathbf{x}_{t-1}),  \label{eq:diffusion_eqn}
\end{align}
where  $0 < \beta_1 <  \beta_2 < \cdots <  \beta_T < 1$  denotes  the ``variance scheduling'' of the Gaussian kernel.   
According to  \eqref{eq:diffusion_eqn}, with $T\!\rightarrow \!\infty$, a data sample ${\bf x}_t$ will   asymptotically  follow  an isotropic Gaussian distribution with  covariance matrix ${\bf \Sigma}\!=\!\sigma^2\mathbf{I}$ for some $\sigma\!>\!0$ \cite{DM_Ho}.   In other words,  the forward diffusion process diminishes  the distinct features of  data samples  gradually.  

In order to run the forward diffusion, the mathematical model in  \eqref{eq:fwd_diffusion} states that  at each   time-step  $t \in [T]$, a new sample should  be drawn from  a conditional Gaussian distribution with  mean vector  ${\mathbf \mu}_t = \sqrt{1 - \beta_t} \mathbf{x}_{t-1}$ and covariance matrix ${\bf \Sigma}^2_t = \beta_t \bf I$. Accordingly, given the variance scheduling $\beta_t$,  the forward diffusion  process can be realized  by sampling from a normal  distribution  $\bm{\epsilon}_{t-1} \sim {\cal N}(\bf 0, I)$  and  diffusing ${\mathbf{x}}_t$ as follows 
\begin{align}\label{eq:fwd_sample_gen_diffusion}
	{\bf x}_t = \sqrt{1-\beta_t}&{\bf x}_{t-1} +\sqrt{\beta_t} {\bm{\epsilon}}_{t-1}.
\end{align}
Exploiting the properties of the summation of two Gaussian random variables,  ${\bf{x}}_t$ at any arbitrary time step $t \in [T]$ can be directly  sampled from $\mathbf{x}_0$. 
This is known as the reparameterization trick in the ML literature \cite{reparam_ML}.     
In other words, \eqref{eq:fwd_sample_gen_diffusion} can be reformulated as 
\begin{align} 
\mathbf{x}_t  &= \sqrt{\bar{\alpha}_t}\mathbf{x}_0 + \sqrt{1 - \bar{\alpha}_t}\bm{\epsilon}_0,  \label{eq:xt_vs_x0} \\ 
    q({\bf x}_t|{\bf x}_0)&\sim\mathcal{N}\left({\bf x}_t;\sqrt{\bar{\alpha}_t}{\bf x}_0,(1-\bar{\alpha}_t)\mathbf{I}\right),\label{eq:xt_vs_x0_dist}
\end{align}
where $\bar{\alpha}_t \overset{\Delta}{=} \prod_{i=1}^t(1-\alpha_i)$ and $\alpha_t  \overset{\Delta}{=}    1-\beta_t$  \cite{reparam_ML}. 

So far, the forward diffusion process has been formulated.  The problem now shifts  to  reversing  the process introduced in \eqref{eq:xt_vs_x0}. Generally speaking, the goal is to generate the desired  samples from  an isotropic  Gaussian noise  
$\mathbf{x}_T$ via sampling from 
$q(\mathbf{x}_{t-1} \vert \mathbf{x}_t)$. 
The challenge is that finding  the conditional probability distribution $q(\mathbf{x}_{t-1} \vert \mathbf{x}_t)$ in an exact closed-form is cumbersome.  This is because for deriving  the aforementioned conditional probability,  we  need  the distribution of all possible data samples. 
Nevertheless, the reverse diffusion process  can be parameterized by ${\bm \theta}$ to  help facilitate  learning  a  probabilistic  model $p_{\bm \theta}(\mathbf{x}_{t-1} \vert \mathbf{x}_t)$ for realizing  the reverse diffusion.    
Based on the above-mentioned facts, we can formulate the parametric  reverse process as 
\begin{align} 
p_{\bm \theta}(\mathbf{x}_{t-1} \vert \mathbf{x}_t) &\sim \mathcal{N}(\mathbf{x}_{t-1}; \boldsymbol{\mu}_{\bm \theta}(\mathbf{x}_t, t), \mathbf{\Sigma}_{\bm \theta}(\mathbf{x}_t, t)),  \label{eq:rev_proc_dist_conditional} 
\\ 
 p_{\bm \theta}(\mathbf{x}_{0:T}) &= p(\mathbf{x}_T) \prod^T_{t=1} p_{\bm \theta}(\mathbf{x}_{t-1} \vert \mathbf{x}_t).  \label{eq:rev_proc_dist_all} 
\end{align}

\begin{figure*} 
\centering
\includegraphics
[width=7.0in,height=2.15in,trim={0.12in 1.5in 1.5in  1.5in},clip]{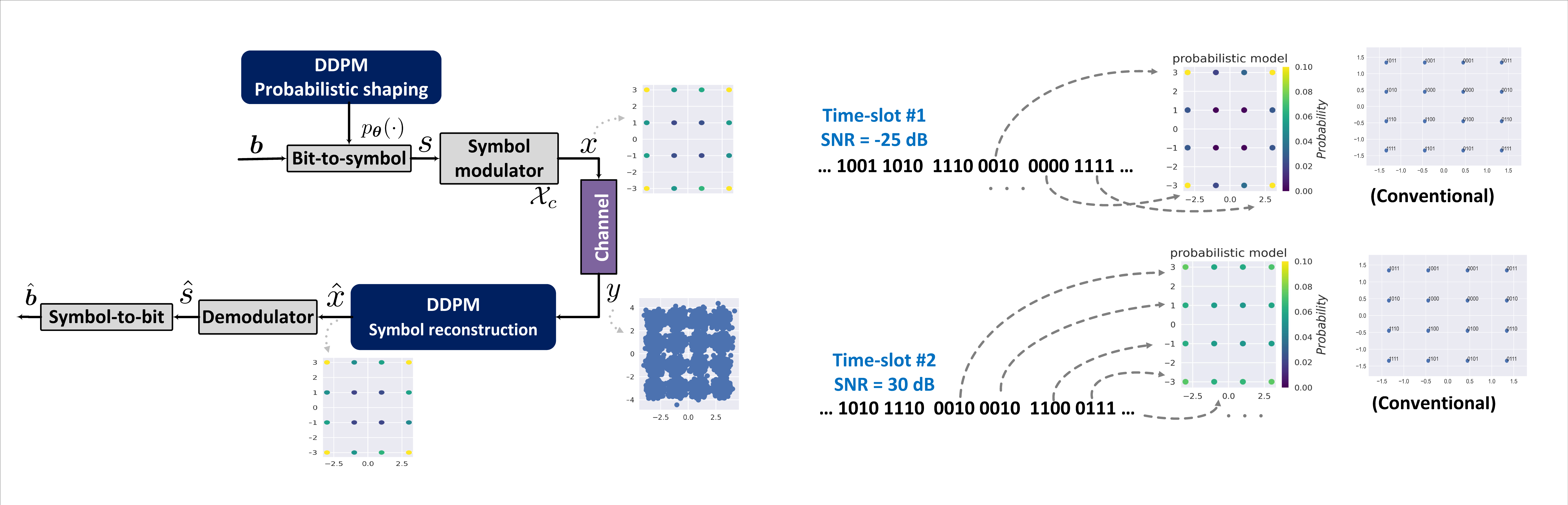}
\vspace{1mm}
\caption{{System model overview. Blocks that are colored in dark blue are of interest in this work.
}}
	\label{fig:SysMod}
 \vspace{-2mm}
\end{figure*}

Invoking \eqref{eq:rev_proc_dist_conditional}, in order to learn the reverse diffusion process,  one needs  to learn  the mean vector  $\boldsymbol{\mu}_{\bm \theta}(x_t,t)$ and the covariance matrix  $\mathbf{\Sigma}_{\bm \theta}(x_t,t)$ 
of  the conditional probability  $p_{\bm \theta}(\mathbf{x}_{t-1} \vert \mathbf{x}_t)$.  Then  a neural network  can be trained for this purpose, and   
${\bm \theta}$ corresponds to  the hyperparameters of that neural network. 
According to \cite{DM_Ho}, we emphasize that the probability distribution of the reverse process could be tractable if conditioned  on ${\bf x}_0$.     
Intuitively speaking,  ${\bf x}_0$ can be interpreted  as a ``reference,'' towards which  we can take  small steps back from noise and generate the data samples. Thus, the conditional  reverse step can be denoted by  $q({\bf x}_{t-1}|{\bf x}_t,{\bf x}_0)$.  
Having  the conditional probabilities of  $q({\bf x}_t|{\bf x}_0)$ and $q({\bf x}_{t-1}|{\bf x}_0)$, one can  utilize Bayes rule to derive the probability distribution of the conditional  reverse step as follows    
\begin{align}
    q(\mathbf{x}_{t-1} \vert \mathbf{x}_t, \mathbf{x}_0) &\sim \mathcal{N}(\mathbf{x}_{t-1}; \hspace{1.5mm} {\tilde{\boldsymbol{\mu}}}(\mathbf{x}_t, \mathbf{x}_0, t), {\tilde{\beta}_t} \mathbf{I}),   \label{eq:rev_conditioned_on_x0}
\end{align}
where we have 
\begin{align}
    {\tilde{\boldsymbol{\mu}}}(\mathbf{x}_t, \mathbf{x}_0, t)
    &=\frac{\sqrt{\alpha_t}(1-\bar{\alpha}_{t-1})}{1-\bar{\alpha}_t}{\bf x}_t + \frac{\sqrt{\bar{\alpha}_{t-1}}
    \beta_t 
    }{1-\bar{\alpha}_t}{\bf x}_0, 
    \label{eq:mu_tilde}  \\ 
   {\tilde{\beta}_t} &=\frac{
   1-\bar{\alpha}_{t-1}}{1-\bar{\alpha}_t} \beta_t.\label{eq:beta_tilde}
\end{align}
As can be observed from \eqref{eq:beta_tilde}, there is no learnable parameter within  the covariance matrix in \eqref{eq:rev_conditioned_on_x0}. 
Therefore, one  should  learn  the  mean vector ${\tilde{\boldsymbol{\mu}}}(\mathbf{x}_t, \mathbf{x}_0, t)$.  
Moreover, applying the  reparameterization trick \cite{reparam_ML} to \eqref{eq:mu_tilde}, 
one  can  rewrite  ${\bf x}_0$  as        
\begin{align}
    \mathbf{x}_0 = \frac{1}{\sqrt{\bar{\alpha}_t}}(\mathbf{x}_t - \sqrt{1 - \bar{\alpha}_t}\bm{\epsilon}_t). \label{eq:x0_vs_xt}
\end{align}
Hence, \eqref{eq:mu_tilde} is  further simplified as follows 
\begin{align}
    \begin{aligned}
\tilde{\boldsymbol{\mu}}(\mathbf{x}_t, \mathbf{x}_0, t) = 
{\frac{1}{\sqrt{\alpha_t}} \Big( \mathbf{x}_t - \frac{1 - \alpha_t}{\sqrt{1 - \bar{\alpha}_t}} \bm{\epsilon}_t \Big)}.
\end{aligned}
\end{align} 
To learn the conditioned probability distribution   $p_{\bm \theta}(\mathbf{x}_{t-1} \vert \mathbf{x}_t)$  
for  the reverse diffusion process, it now suffices to  train  a neural network with the objective of  approximating  $\tilde{\boldsymbol{\mu}}(\mathbf{x}_t, \mathbf{x}_0, t)$.     
To do so, one can formulate  the approximated mean vector  $\boldsymbol{\mu}_{\bm \theta}(\mathbf{x}_t, t)$  with the same mathematical form  as the ground-truth mean vector $\tilde{\boldsymbol{\mu}}(\mathbf{x}_t, \mathbf{x}_0, t)$.  
Moreover, considering the fact that at time-step $t$,  $\mathbf{x}_t$  is known,  
the neural network can equivalently  approximate $\bm{\epsilon}_t$.  
Therefore, $\boldsymbol{\mu}_{\bm \theta}(\mathbf{x}_t, t)$ can be expressed as   
 \begin{align}\label{eq:mu_theta_to_learn}
\boldsymbol{\mu}_{\bm \theta}(\mathbf{x}_t, t) &=  {\frac{1}{\sqrt{\alpha_t}} \Big( \mathbf{x}_t - \frac{1 - \alpha_t}{\sqrt{1 - \bar{\alpha}_t}} \bm{\epsilon}_{\bm \theta}(\mathbf{x}_t, t) \Big)},  
\end{align}
where   $\bm{\epsilon}_{\bm \theta}(\mathbf{x}_t, t)$ represents the output of the  neural network.    
Now the goal is to minimize the difference between $\boldsymbol{\mu}_{\bm \theta}(\mathbf{x}_t, t)$ and  $\tilde{\boldsymbol{\mu}}(\mathbf{x}_t, \mathbf{x}_0, t)$.    Hence, the corresponding  loss function $\mathcal{L}_t$  ($\forall t \in [T]$)  for the reverse diffusion framework can be formulated as  
\begin{align}
\mathcal{L}_t &=  
{\mathbb{E}}_{\begin{subarray}{l}t\sim {\mathsf{Unif}}[T]\\ {\bf x}_0\sim q({\bf x}_0) \\ \bm{\epsilon}_0\sim \mathcal{N}(\mathbf{0},\bf{I})\\ \end{subarray}}
\Big[\|\bm{\epsilon}_t - \bm{\epsilon}_{\bm \theta}(\mathbf{x}_t, t)\|^2 \Big]  \nonumber \\  
&=  {\mathbb{E}}_{\begin{subarray}{l}t\sim {\mathsf{Unif}}[T]\\ {\bf x}_0\sim q({\bf x}_0) \\ \bm{\epsilon}_0\sim \mathcal{N}(0,\bf{I})\\ \end{subarray}}  \Big[\|\bm{\epsilon}_t - \bm{\epsilon}_{\bm \theta}(\sqrt{\bar{\alpha}_t}\mathbf{x}_0 + \sqrt{1 - \bar{\alpha}_t}\bm{\epsilon}_t, t)\|^2 \Big], \label{eq:loss_func}
\end{align}
where  $\bm{\epsilon}_t$ denotes the  diffused noise   at time step $t$, and $\bm{\epsilon}_{\bm \theta}(\mathbf{x}_t, t)$ denotes  the approximated noise vector  at the  output of the  neural network.    
The  diffusion model is trained based on the loss function given in \eqref{eq:loss_func},  using an error measure, e.g., the widely-used MSE, between the injected  and the predicted noise, where in our case,  $\mathbf{x}_0$  stands for  data samples from the constellation symbols in $\mathcal{X}_c$.  
{Invoking \eqref{eq:loss_func},  we  also emphasize that $\bar{\alpha}_t$, $\forall t\in [T]$,  is a function of the  variance scheduling $\beta_t$  
that can help design different loss functions 
$\mathcal{L}_t$ during training.  
This leads to the fact that different noise levels would be seen during training, which can  make the system robust against a wide range of noise levels during the sampling phase as well.}

\section{System Model and Proposed Scheme}\label{sec:SysMod}
In this section, we first present our system model.  
Then, our  DDPM-based solution and the corresponding algorithms for probabilistic constellation shaping are proposed.

\subsection{System Model}
Fig. \ref{fig:SysMod} demonstrates the communication  system model considered in this paper.  
The system takes the information bit-stream and maps it onto hypersymbols $s \in \mathcal{S}$ according to the learnable  distribution $p_{\bm \theta}(s)$  (parameterized by $\bm \theta$), where  $\mathcal{S}  = \{1,\dots,M\}$  denotes the set of all hypersymbols, and  $M$ denotes the modulation order.   
In this paper, the probabilistic constellation shaping $p_{\bm \theta}(s)$ 
is realized by a DDPM, which is trained and  employed  at  the transmitter and the receiver. 
Details on  the training and sampling processes of the employed DDPM are elaborated in the next subsection. 
The sequence of hypersymbols is then fed into a symbol modulator which maps each symbol $s$ into a constellation point $x \in \mathcal{X}_c$, with ${\cal X}_c$ showing the set of constellation points. 
Each symbol is  generated  according to the distribution $p_{\bm \theta}(s)$.    
In other words, the frequency of sending a bit-stream over the constellation point $x = g(s)$  corresponds to the parametric distribution $p_{\bm \theta}(s)$, where $g$ denotes  the  modulation functionality.  Accordingly,    
the distribution of $x$, $\forall x \in \mathcal{X}_c$,  can be written as 
\begin{align}\label{eq:pX}
p_{\bm \theta}(x) = \sum_{s \in \mathcal{S}} \delta\left(x - g(s)\right)p_{\bm \theta}(s). 
\end{align}
As explained in Section \ref{sec:Intro}, our focus in this paper  is on probabilistic  constellation shaping, and hence, the   geometry of constellation is determined in advance according to network  configurations.\footnote{We adhere to  standard constellation schemes, such as quadrature amplitude modulation (QAM), in order to propose a system which is compliant with the real-world communication systems.}   
Thus, the design of modulator function $g(\cdot)$ is not  of interest in this work.\footnote{Mathematically speaking, the modulator can be considered as  a matrix $\mathbf{C} {=} [\mathbf{c}_1,\ldots,\mathbf{c}_i,\cdots,\mathbf{c}_{2^M}]^{\sf T} \in 2^M \times 2$,  
with rows  $\mathbf{c}^{\mathsf{T}}_i \in \mathbb{R}^2,~ i\in [2^M]$  showing the 
constellation point locations.  
One can take the product of $\mathbf{C}$ and  a one-hot vector $\mathbf{s}$, with the $s$-th element set to one,  to select  a constellation point corresponding to $s$.} 
Accordingly, we have a one-to-one mapping between the constellation point $x$
and the information  symbol $s$, and one can consider that  the transmitter's output  is directly   sampled from  $p_{\bm \theta}(\cdot)$.  
In addition, similar to \cite{constell}, we assume that the bit-to-symbol mapper is  known.     
Information-bearing signal $x$ is then sent over the communication  channel, and the channel output $y$ is observed at the receiver. 
Then the receiver needs to reconstruct the transmitted symbols by approximating the posterior distribution $p(s|y)$
given the channel output.     
To do so, 
the receiver leverages the  DDPM,   
parameterized by $\bm \theta$ with the same architecture as that of the transmitter,    
and  maps each received sample $y$ to a probability distribution  over the set of hypersymbols $\cal S$.  
Having this approximation,  symbols can  be  reconstructed   at the receiver's  demodulator ($\hat{s}$ in Fig. \ref{fig:SysMod}), and  
the information bits 
can be obtained  
using the typical  symbol-to-bit  mappers.

In the next subsection, we address our proposed scheme and  provide the details on how to leverage DDPMs for the considered system model.

\begin{figure}
\centering
    \begin{subfigure}{0.5\textwidth}
        \centering
        \includegraphics[width=\linewidth, trim={3.0in 0.0in 2.0in  0.0in},clip]{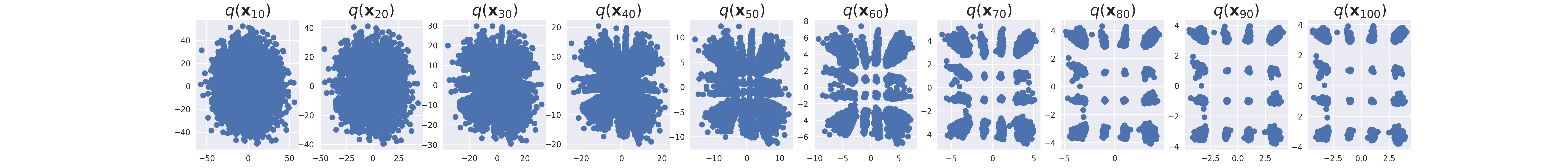}
    \end{subfigure}
    \begin{subfigure}{0.4\textwidth}
        \centering
        \includegraphics[width=\linewidth, trim={8.7in 0.0in 0.8in  0.0in},clip]{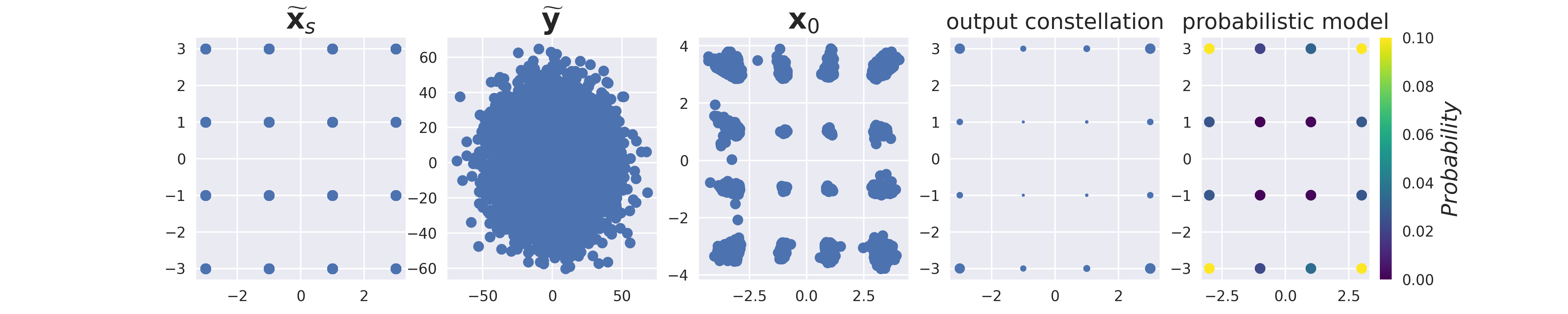} 
    \end{subfigure}
        \begin{subfigure}{0.5\textwidth}
        \centering
        \includegraphics[width=\linewidth, trim={3.0in 0.0in 2.0in  0.0in},clip]{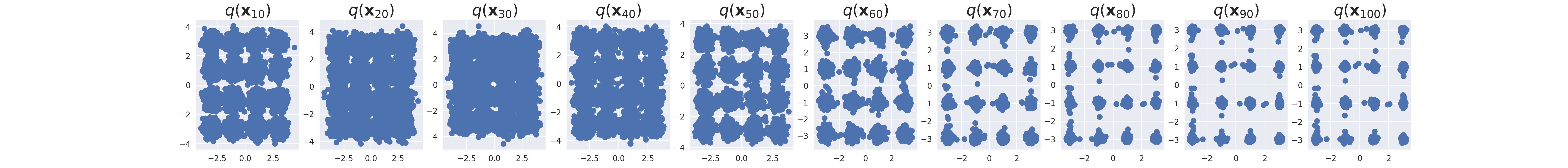}
    \end{subfigure}
    \begin{subfigure}{0.4\textwidth}
        \centering
        \includegraphics[width=\linewidth, trim={8.7in 0.0in 0.8in  0.0in},clip]{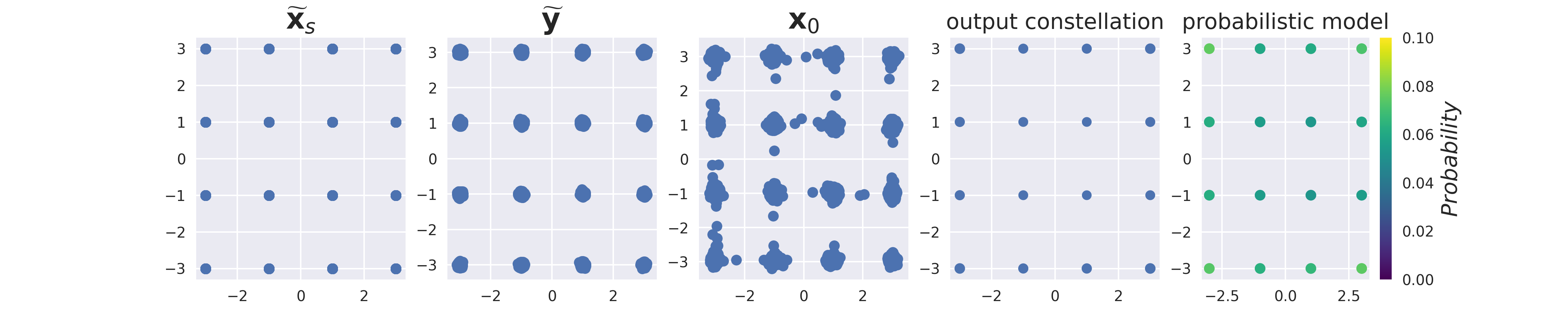} 
    \end{subfigure} 
    \caption{{A visual insight on the process of  constellation shaping via our proposed scheme. From top, denoising and generation process of the diffusion model, and the  probabilistic constellation shaping at the transmitter are shown, respectively. This is carried out  for two different SNR values of  $-25$ dB  and $30$ dB, respectively.}} 
    \label{fig:insight}
    \vspace{-2mm}
\end{figure}

\subsection{Proposed Approach}\label{subsec:proposed_approach}   
{
We aim to probabilistically  shape the constellation symbols by finding a proper $p_{\bm \theta}(\cdot)$, such that  
the information-bearing symbols 
sent by 
the transmitter,  
and what is inferred (reconstructed) at the receiver 
become as similar as possible,\footnote{This ``similarity'' will be quantitatively evaluated in the next section, using two widely-adopted metrics of mutual information and cosine similarity.} 
resulting in as few mismatches between the communication parties as possible.  
This fact, together with the characteristic of  diffusion models to ``denoise-and-generate'',  motivates us to propose a  DDPM-based solution  for probabilistic constellation shaping.    
The key idea to fulfill the desired  similarity  is that  
the transmitter ``mimics'' the way the receiver would perform   the reconstruction  
of symbols out of noisy signals.  
For instance, when the communication channel is experiencing high levels of noise, i.e.,  in low-SNR regime,  we intuitively expect that  most often, the  receiver would be able to  decode the symbols corresponding to the 
points that are relatively far from each other in the constellation geometry, while the other points are prone to being decoded incorrectly.  
Thus, the transmitter  could  assign the information bits to those constellation points that are at the furthest distance from each other, with higher probabilities.    
This also helps facilitate having 
``mutual understanding'' of how to map and de-map the information symbols over time, realizing \emph{native intelligence} among communication parties.}  
{To gain a visual insight on the process of  constellation shaping via our proposed scheme, Fig. \ref{fig:insight}  demonstrates the shaped constellation points and their corresponding probability distribution for 16-QAM geometry for two different  SNR values of $-25$ dB and $30$ dB.  
More details are elaborated on  in Section \ref{sec:Eval}. 
}

Motivated by the abovementioned  discussions,  our step-by-step solution  can be elaborated on as follows.    
\subsubsection{DDPM Training}\label{step1} 
A  DDPM  
 is trained based on the loss function given in \eqref{eq:loss_func}. This  corresponds to training the parameter $\bm \theta$ for our probabilistic shaping scheme  in \eqref{eq:pX}.    
The  process 
is summarized in Algorithm \ref{alg:trainAlg}.    
We take a random sample $\mathbf{x}_0$ from the set  of constellation points $\mathcal{X}_c$;  
We also  sample a random time-step $t \sim \mathsf{Unif}[T]$.   
A noise vector $\boldsymbol{\epsilon}$ (with the same shape as the input) is also sampled from  the normal  distribution.  
Given the sampled time-step $t$, the  noise level is adjusted according to the  
variance scheduling  $\beta_t$. We then make the input vector $\mathbf{x}_0$ noisy according to \eqref{eq:xt_vs_x0}.   
The neural network is trained based on  \eqref{eq:loss_func} to approximate  the noise vector lying in the noisy data.   
{Training can be carried out in a central cloud, or an edge server,   
and then,  the trained model is downloaded by the communication entities.}    
{Note that the same trained model  
is  deployed at both the  transmitter and the receiver, and  they use it to run their DDPM-based  framework for the constellation generation and decoding, respectively.    
}

\begin{figure}[t]
\vspace{-2mm}
\begin{algorithm}[H]
\hspace*{0.02in} {\bf {Hyper-parameters:}}
	{Number of time-steps $T$, neural architecture $\boldsymbol{\epsilon}_{\bm \theta}(\cdot, t)$, variance schedule   $\beta_t$, and $\bar{\alpha}_t, \forall t \in [T]$.} \\
    \hspace*{0.02in} {\bf {Input:}}
	{Sample points from  the constellation geometry $\mathcal{X}_c$.} \\
	\hspace*{0.02in} {\bf {Output:}} {Trained neural model for DDPM.}
	\caption{
 \small
 Training algorithm of DDPM \cite{DM_Ho} }
	\label{alg:trainAlg}
	\begin{algorithmic}[1] 
    \WHILE {the stopping criteria are not met}
    \STATE Randomly sample $\mathbf{x}_0$ from  $\mathcal{X}_c$
    \STATE Randomly sample $t$ from $\mathsf{Unif}[T]$ 
    \STATE Randomly sample $\boldsymbol{\epsilon}$ from $\mathcal{N}(\mathbf{0},\mathbf{I})$ 
      \STATE Take gradient descent step on
      \STATE $\qquad {\grad}_{\boldsymbol{\theta}} \left\| \bepsilon - \bepsilon_{\bm \theta}(\sqrt{\bar\alpha_t} \bx_0 + \sqrt{1-\bar\alpha_t}\bepsilon, t) \right\|^2$
    \ENDWHILE
	\end{algorithmic}
\end{algorithm}  
\vspace{-5mm}
\end{figure}

\subsubsection{Link quality estimation with channel SNR}
Within each transmission slot,  the transmitter  first  estimates the quality of the communication link. This is done  using the pilot signal sent by the destination node  at the beginning of each transmission slot, and the SNR level of communication channel can be calculated 
\cite{Derrick_MIMO}.

\begin{figure*}
	\vspace{0mm}
	\centering
	\includegraphics
	[width=6.5in,height=1.2in,trim={0.0in 0.0in 0.0in  0.0in},clip]{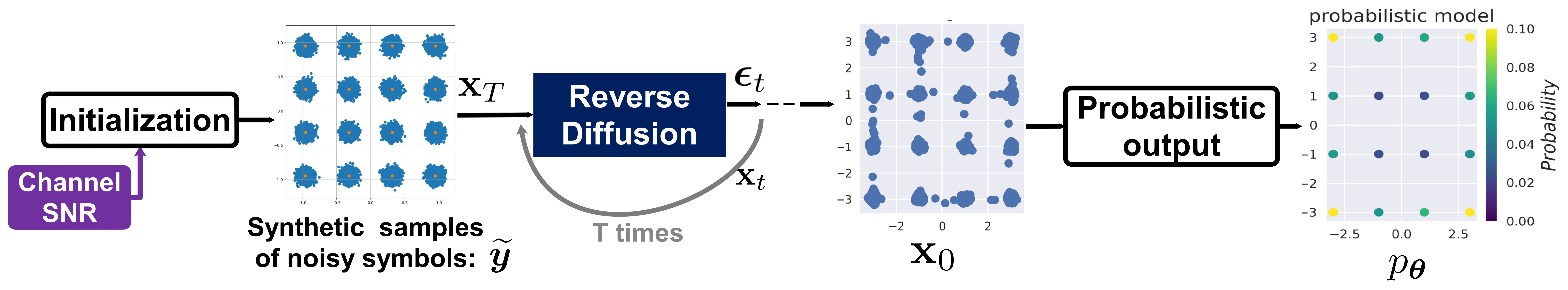}
	\vspace{3mm}\caption{{Block-diagram of Algorithm \ref{alg:sampling_Tx} for the proposed scheme.}} 
	\label{fig:Step3_prob_constell}
 \vspace{0mm}
\end{figure*}

 \begin{figure}[t]
\vspace{-2mm}
\begin{algorithm}[H]
\hspace*{0.02in} {\bf {Hyper-parameters:}}
	{Number of time-steps $T$, trained neural model $\bm \theta$, constellation geometry $\mathcal{X}_c$} \\
    \hspace*{0.02in} {\bf {Input:}}
	{Channel SNR $\Gamma$.} 
  \caption{ \small
  DDPM sampling: Probabilistic shaping at transmitter}  \label{alg:sampling_Tx}
\begin{algorithmic}[1]
    \vspace{0.0in}  
\STATEx \hspace{-1.5em} {\bf{Initialization: {Synthetic samples of noisy symbols:}}} 
    \STATE Randomly sample $\widetilde{\mathbf{x}}_s$, with size $N_s$, from set  $\mathcal{X}_c$ 
    \STATE  Randomly sample $\widetilde{\mathbf{n}}$, with size $N_s$,  from $\mathcal{N}(\mathbf{0}, \mathbf{I})$ 
    \STATE $\widetilde{\boldsymbol{y}} = \widetilde{\mathbf{x}}_s + \delta  \widetilde{\mathbf{n}}$ 
\STATEx \hspace{-1.5em} {\bf Reverse diffusion:}    
    \STATE  $\bx_T = \widetilde{\boldsymbol{y}}$  
 \FOR {$t=T, ... , 1$}
      \STATE  $\bz \sim \mathcal{N}(\bzero, \bI)$ if $t > 1$, else $\bz = \bzero$
      \STATE  $\bx_{t-1} = \frac{1}{\sqrt{\alpha_t}}\left( \bx_t - \frac{1-\alpha_t}{\sqrt{1-\bar\alpha_t}} \bepsilon_{\bm \theta}(\bx_t, t) \right) + \sqrt{1-\alpha_t} \bz$
   \ENDFOR 
\STATEx \hspace{-1.5em} {\bf Probabilistic output:}   
    \STATE  $\boldsymbol{\psi} = \mathtt{proj}_{\mathcal{X}_c}(\mathbf{x}_0)$
    \STATE  $\boldsymbol{c} \hspace{1mm} =  \mathtt{count} \left(\boldsymbol{\psi}, \mathcal{X}_c\right)$  
    \STATE  \textbf{return} $p_{\bm \theta} = 
     {\boldsymbol{c}}/{N_s}$
    \vspace{0.0in}
  \end{algorithmic}
\end{algorithm}
\vspace{-5mm}
\end{figure}

\subsubsection{\textbf{Probabilistic shaping at the transmitter}} 
{The DDPM is run at the transmitter to probabilistically shape (generate) the constellation symbols according to the channel SNR. 
To fulfill the desired ``similarity'' between the  transmitter and the receiver, 
the transmitter synthetically generates samples of the received signal (with the SNR level according to the estimated channel SNR), and then tries to denoise   them to infer  which constellation points are more probable to be reconstructed, sending the information bits over those points.}    
To do so, we  first  
take $N_s$ samples 
from the set of constellation symbols $\mathcal{X}_c$ uniformly at random. {The sample size $N_s$ can be regarded as the  number of observations  
to form (generate) the empirical distribution of our probabilistic shaping.}    
In addition,  we  sample $N_s$ realizations of  random 
noise with    
average power $\delta^2$, and inject them  into the uniformly-sampled symbols.  
The power of synthetic noise, $\delta^2$, is calculated according  to the channel SNR, $\Gamma$,  which was obtained at Step 2. This  can be formulated  as 
\begin{align}\label{eq:noise_power}
\delta ^ 2 = {10^{{ \Gamma}/{10}}} \mathsf{P} ,
\end{align} 
where $\mathsf{P}$ denotes the average transmit power. 
The 
noisy 
version of samples is then fed into the trained DDPM,  and   the  reverse diffusion process is run to denoise and  generate  symbols out of the synthetically-noisy samples.  
The distribution of the generated  samples at the output of the DDPM block is considered 
as the output  probabilistic  constellations,  onto which the information symbols are mapped to be sent.

The overall algorithm is proposed in Algorithm \ref{alg:sampling_Tx}.  {Moreover, a block-diagram of the proposed probabilistic constellation shaping algorithm  is illustrated in Fig. \ref{fig:Step3_prob_constell}.}  
{In Algorithm \ref{alg:sampling_Tx},
Lines 2 to 3 correspond to the  synthetic generation of noisy received signals, given the channel SNR.   
Moreover, the main loop   corresponds to the reverse diffusion process from time-step $T$ to $1$, according to \ref{eq:mu_theta_to_learn}, which tries to mimic the denoising functionality of the receiver, using the trained  DDPM.}      
Also, $\mathtt{proj}_{\mathcal{S}}(\mathbf{x})$ stands for the projection operator, which  maps  the elements  of vector $\mathbf{x}$ onto the nearest elements in the set $\mathcal{S}$.  Moreover,  $\mathtt{count}(\mathbf{x}, \mathcal{S})$ outputs a vector with 
size $|\mathcal{S}|$,  
with elements representing  the number of occurrences of the elements of set $\cal S$ in vector $\mathbf{x}$.      
Notably,  $\boldsymbol{\psi}$ in Algorithm \ref{alg:sampling_Tx}  denotes the probabilistically-shaped constellation points at the output of the transmitter's DDPM block, and $p_{\bm \theta}$ stands for the corresponding distribution inferred by the diffusion model.  {In the next section, data visualization of different steps are presented with numerical examples, to further  understand what is going on in the proposed constellation shaping algorithm.}

\begin{figure*}[] 
\centering
\includegraphics
[width=6.75in,height=1.75in,trim={0.0in 0.0in 0.0in  0.0in},clip]{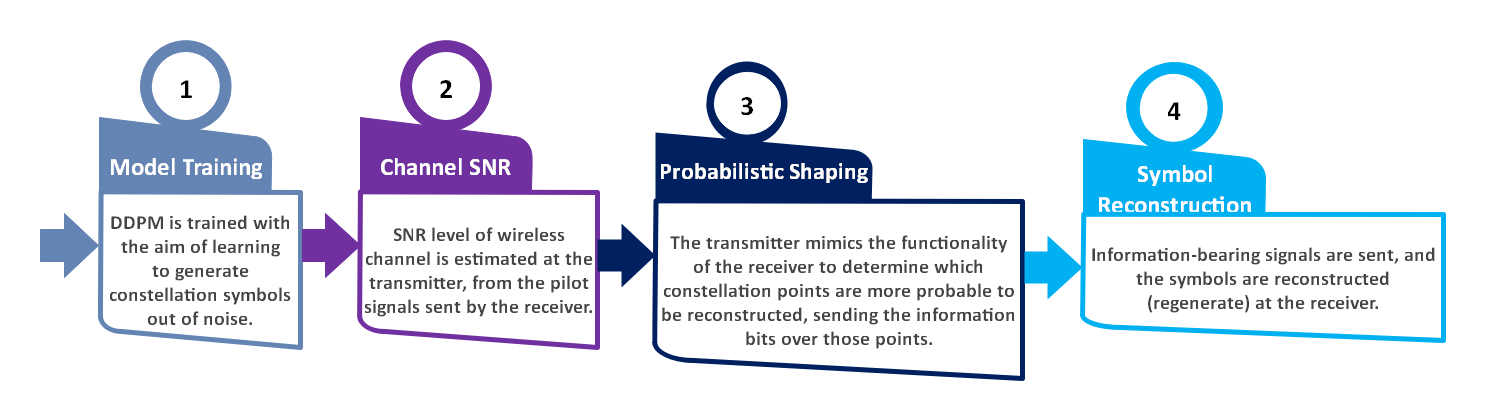}
\vspace{2mm}
\caption{{Summary of the  proposed approach.  
}}
	\label{fig:diagram}
 \vspace{-2mm}
\end{figure*}

\subsubsection{Symbol reconstruction at the receiver} 
After  generating constellation symbols, 
 information signals are transmitted  according to the probabilistic model of the constellation points. The symbols are then   received by the receiver, and    
it runs the diffusion model 
to reconstruct (regenerate) the symbols from the noisy  received  signals.   The corresponding  algorithm for this step is proposed in Algorithm \ref{alg:sampling_Rx}. 
Starting from the received batch of noisy symbols, denoted by $\boldsymbol{y}$, for each time step $t\in\{T, T-1, \ldots, 1\}$,  the neural network  outputs $\bm{\epsilon_{\bm \theta}}(\hat{\mathbf{x}}_t, t)$ to approximate  the residual noise within the batch of symbols, and the sampling algorithm is run according to Line $4$ of the algorithm, in order  to sample $\hat{\mathbf{x}}_{t-1}$. The process is executed for $T$ steps.\footnote{We emphasize that within each transmission slot, while the channel coherence time is respected,  the channel SNR remains unchanged 
compared to the one that is utilized by the transmitter for the  constellation shaping.}       

To summarize  Section \ref{subsec:proposed_approach},  a general overview of the proposed scheme  is illustrated  in Fig. \ref{fig:diagram}.

\begin{figure}[t]
\vspace{-2mm}
\begin{algorithm}[H]
\hspace*{0.02in} {\bf {Hyper-parameters:}}
	{Number of time-steps $T$, trained neural model $\bm \theta$, constellation geometry $\mathcal{X}_c$} \\
    \hspace*{0.02in} {\bf {Input:}}
	{Received signal $\boldsymbol{y}$} 
  \caption{
  \small  
  DDPM sampling: Symbol reconstruction at receiver}  \label{alg:sampling_Rx}
  \begin{algorithmic}[1]
    \vspace{0.0in}
    \STATE $\hat{\mathbf{x}}_T = \boldsymbol{y}$ 
    \FOR{$t=T, ... , 1$}
      \STATE $\bz \sim \mathcal{N}(\bzero, \bI)$ if $t > 1$, else $\bz = \bzero$
      \STATE $\hat{\mathbf{x}}_{t-1} = \frac{1}{\sqrt{\alpha_t}}\left( \hat{\mathbf{x}}_t - \frac{1-\alpha_t}{\sqrt{1-\bar\alpha_t}} \bepsilon_{\bm \theta}(\hat{\mathbf{x}}_t, t) \right) + \sqrt{1-\alpha_t} \bz$
    \ENDFOR
    \STATE \textbf{return} $\mathtt{proj}_{\mathcal{X}_c}(\hat{\mathbf{x}}_0)$  
  \end{algorithmic}
\end{algorithm}
 \vspace{-8mm}
\end{figure}

\begin{figure*}
	\vspace{0mm}
	\centering
	\includegraphics
	[width=5.2in,height=2.0in,trim={0.0in 0.0in 0.0in  0.0in},clip]{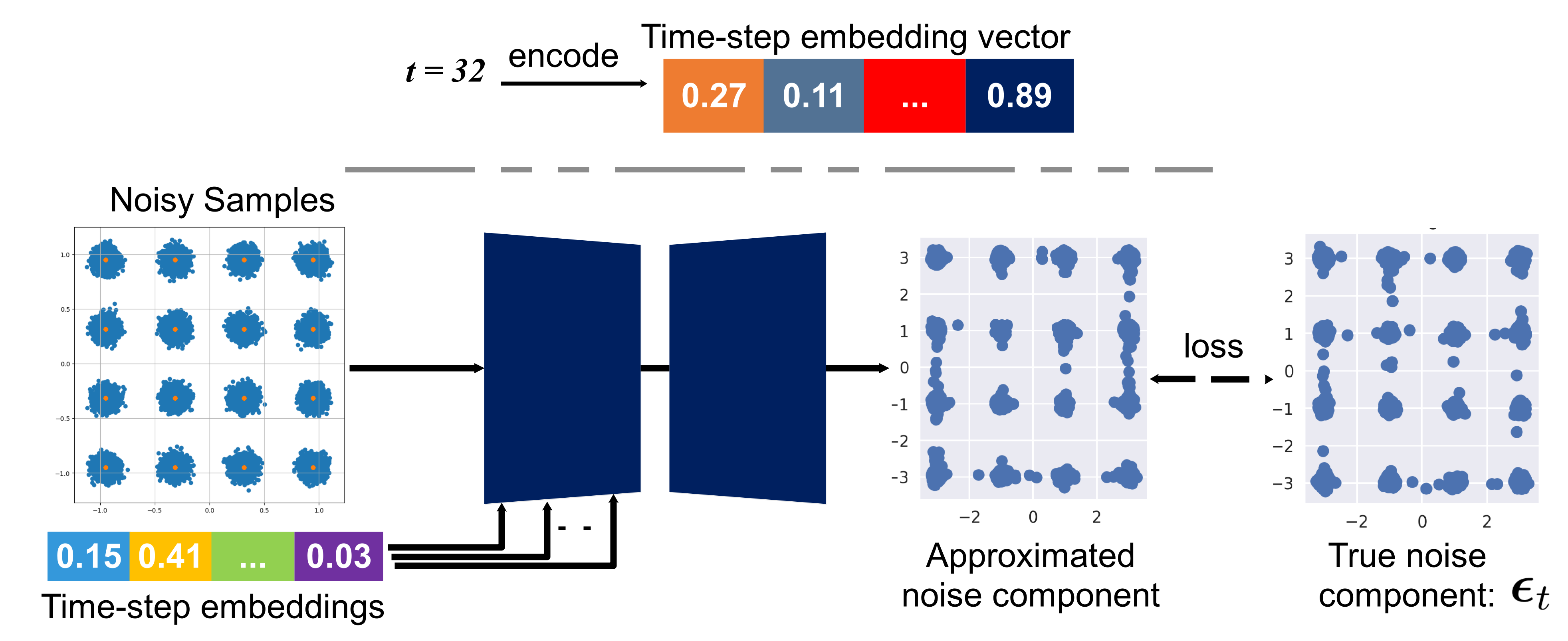}
	\vspace{3mm}\caption{{Block  diagram of training the diffusion model. Time-steps are incorporated into the hidden layers as embedding vectors.}} 
	\label{fig:embedding_constell}
 \vspace{0mm}
\end{figure*}

\section{Numerical Experiments}\label{sec:Eval}
In this section, we carry out different  numerical evaluations  in terms of  mutual information metric and cosine similarity, in order  to highlight the performance of the proposed scheme compared to other benchmarks.  We show that our DDPM-based approach  achieves  $30\%$ improvement in terms of cosine similarity,  and a threefold improvement in terms of mutual information metric compared to DNN-based approach.   
We also show that our proposed scheme maintains  \emph{native resilience} as well as \emph{robust out-of-distribution
performance} under  low-SNR regime  and non-Gaussian assumptions.

\begin{figure} 
\centering
\includegraphics
[width=3.45in,height=2.55in,trim={0.3in 0.1in 0.5in  0.4in},clip]
{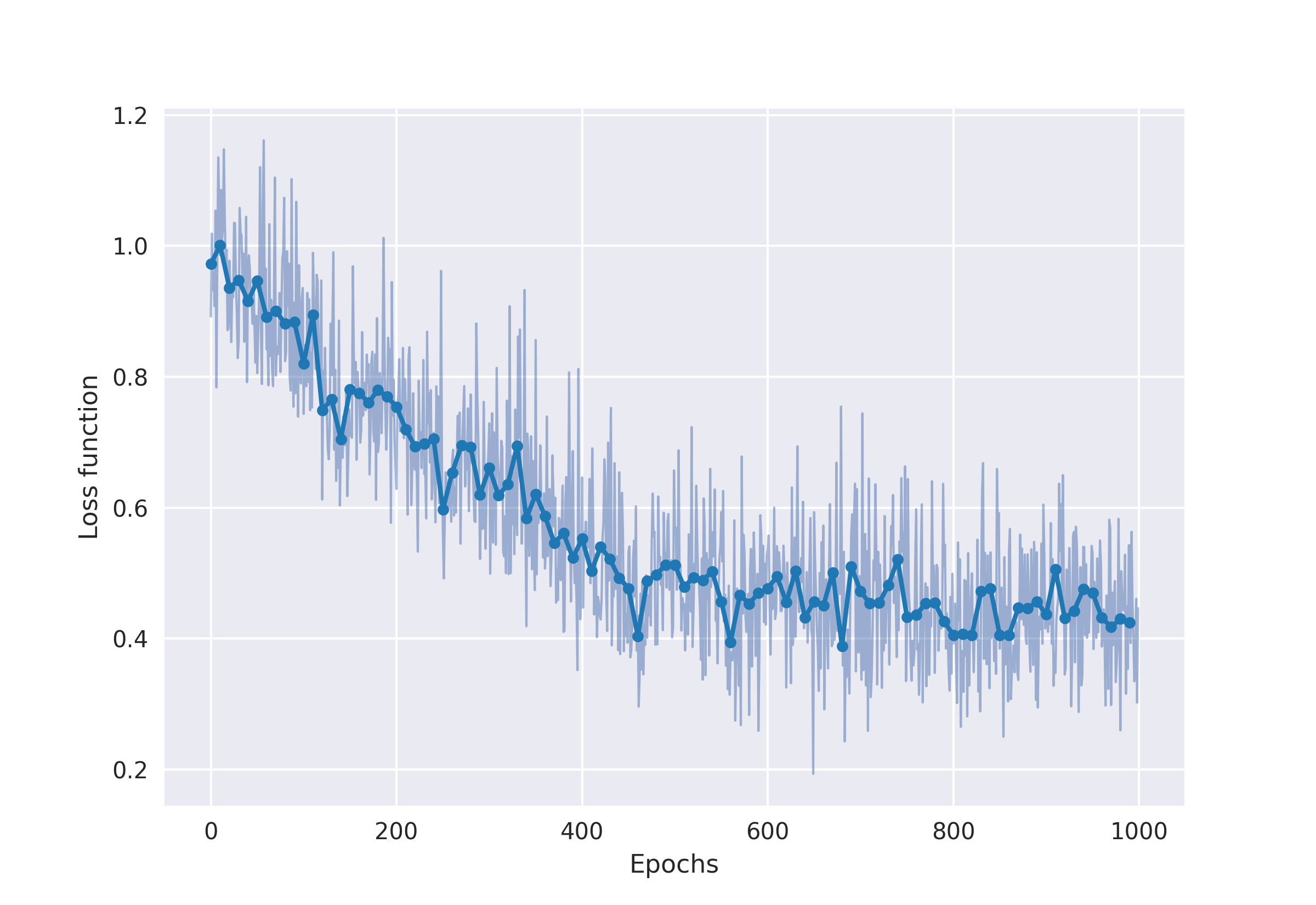} 
\vspace{-2mm}
\caption{ 
Training process of DDPM over epochs.}
	\label{fig:TP}
 \vspace{-3mm}
\end{figure}

To parameterize the reverse diffusion process of our DDPM-based scheme, we employ a neural network 
 comprised of 
$3$ hidden conditional layers (with softplus activation functions), each with $128$
neurons conditioned on $t$.   
The output layer is a  linear layer with the same size as the input.    
Instead of training $T$ distinct models for each time-step, we employ only  one neural  model for  the entire denoising time-steps. This is done via  sharing the parameters of the neural network across time-steps---we encode the time-steps $t \in [T]$ and input it to each hidden layer of our neural network  as a 
vector embedding, {as illustrated in Fig. \ref{fig:embedding_constell}.}    
The hidden layers are conditioned on $t$  via being   multiplied by the time embeddings \cite{CGM_ChanEst_WCNC, DM_Ho, DM_survey}.   
Intuitively speaking, incorporating  the time-step into the neural network makes our model  ``know''  at which particular time-step it is operating for every sample.  
To stabilize the training algorithm of our DDPM framework,   exponential moving average (EMA) method  is  implemented \cite{EMA}. To elaborate, instead of directly updating the weights, a copy of the previous weights is kept, and the weighted average  between the previous and new version of the weights are calculated  for the model update.   
This  helps maintain model momentum.  
For the diffusion process, the variance scheduling,   $\alpha_t$ is set to constants decreasing  from $\alpha_1=0.99999$ to $\alpha_T=0.99$ with Sigmoid scheduling \cite{DM_for_E2EComm}. 
For training the diffusion model, we use adaptive moment estimation (Adam) optimizer with  learning rate  $\lambda = 10^{-3}$. We consider QAM geometry as a   widely-adopted constellation format in wireless networks \cite{twelve_6G, hexaX, constell}.  Moreover, we set  $T=100$ and $T = 200$ for 16-QAM and 64-QAM geometry, respectively.  The stopping criterion in Algorithm \ref{alg:trainAlg} is met when reaching the maximum number of epochs \cite{CDDM, CGM_ChanEst_WCNC}, which is set to $1000$ epochs for 16-QAM and $5000$ epochs for 64-QAM.    
The training process for the case of 16-QAM geometry is illustrated in Fig. \ref{fig:TP}. We can observe a decreasing trend in the training loss function over epochs.  The same trend could be observed for 64-QAM case, and   we do not repeat the figure due to space limitations.

\begin{figure}
\centering
    \begin{subfigure}{0.5\textwidth}
        \centering
        \includegraphics[width=\linewidth, trim={5.35in 0.0in 2.0in  0.0in},clip]{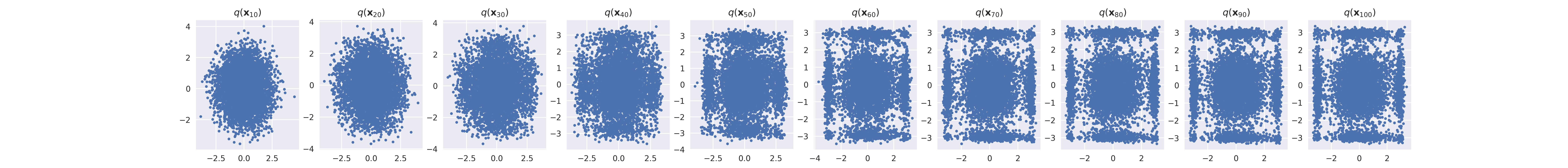}
    \end{subfigure}
    \begin{subfigure}{0.5\textwidth}
        \centering    
        \includegraphics[width=\linewidth, trim={5.35in 0.0in 2.0in  0.0in},clip]{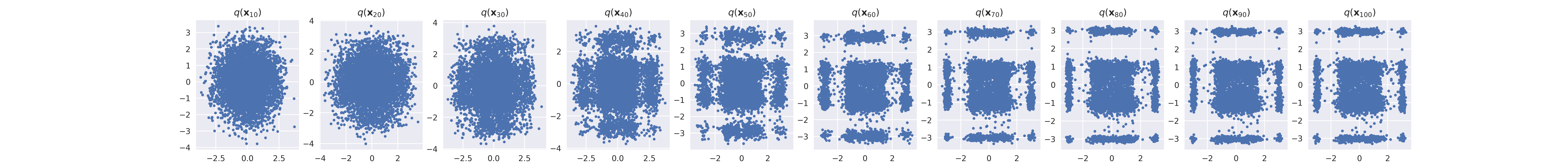}
    \end{subfigure}
    \begin{subfigure}{0.5\textwidth}
        \centering    
        \includegraphics[width=\linewidth, trim={5.35in 0.0in 2.0in  0.0in},clip]{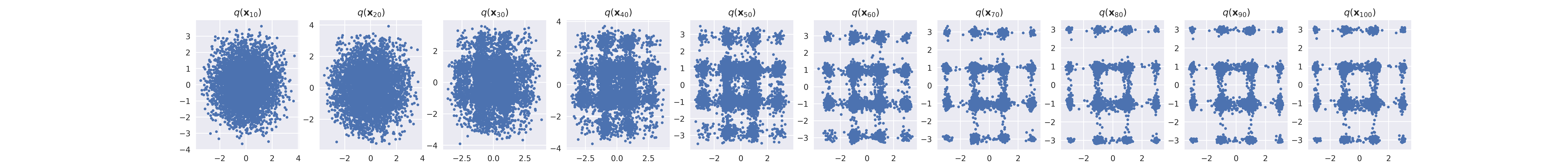}
    \end{subfigure}
    \begin{subfigure}{0.5\textwidth}
        \centering    
        \includegraphics[width=\linewidth, trim={5.35in 0.0in 2.0in  0.0in},clip]{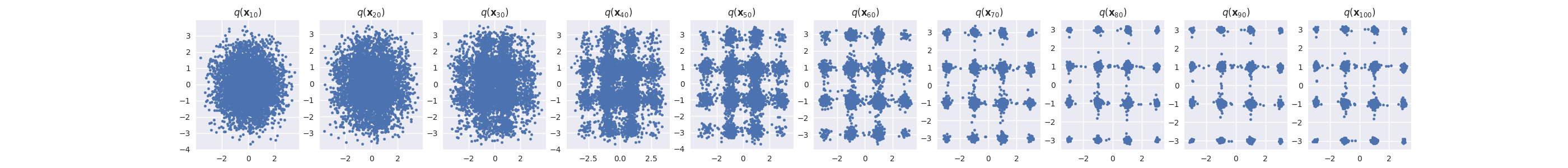}
    \end{subfigure}
    \caption{Data visualization for the reverse diffusion process during training (16-QAM).  
    From top to bottom, the rows correspond to epochs  
    $200$, $300$, $400$,  
   and  $1000$, respectively.  }
    \label{fig:train_16qam}
    \vspace{-3mm}
\end{figure}

\begin{figure}
\centering
    \begin{subfigure}{0.5\textwidth}
        \centering    
        \includegraphics[width=\linewidth, trim={5.35in 0.0in 2.0in  0.0in},clip]{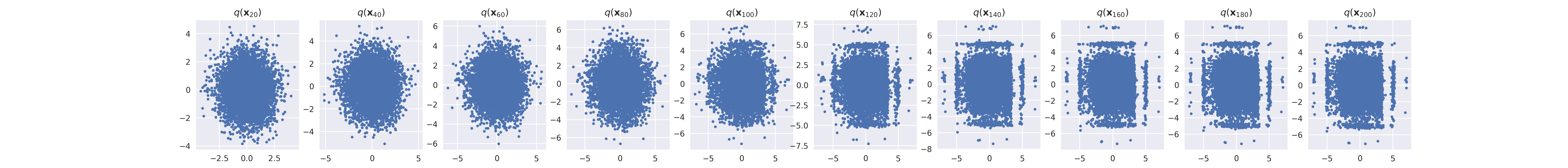}
    \end{subfigure}
    \begin{subfigure}{0.5\textwidth}
        \centering    
        \includegraphics[width=\linewidth, trim={5.35in 0.0in 2.0in  0.0in},clip]{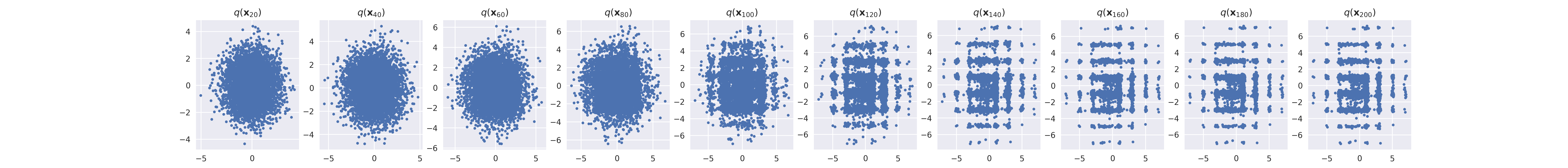}
    \end{subfigure}
    \begin{subfigure}{0.5\textwidth}
        \centering    
        \includegraphics[width=\linewidth, trim={5.35in 0.0in 2.0in  0.0in},clip]{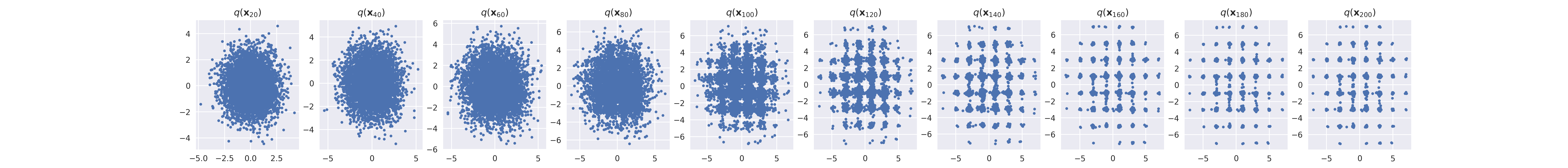}
    \end{subfigure}
    \caption{Data visualization for the reverse diffusion process during training (64-QAM).  
   From top to bottom, the rows correspond to epochs 
   $1500$, $2500$,  
  and  $4500$, respectively. 
 }
    \label{fig:train_64qam}
    \vspace{-3mm}
\end{figure} 

\subsection{Data Visualization of The Proposed  Approach}\label{ssec:evaluations_train}
In this subsection, we  carry out {data visualization} for the generation process  of our DDPM-based solution  during training and sampling.

Figs. \ref{fig:train_16qam}  and \ref{fig:train_64qam} visualize the reverse diffusion  process across epochs during the training of the implemented  DDPM.      
For these figures, we take ``snapshots" of the model  (i.e., we save the model's current state)  at specific checkpoints (epochs) during the training process,  
and plot  the output of the DDPM block over time-steps.
As can be seen from the figures, the employed  DDPM gradually learns to  generate the desired  samples (constellation symbols) out of an isotropic Gaussian noise. Moreover, we can  observe from the figures that as we reach the maximum number of  epochs, the model can sooner (i.e., in fewer time-steps) generate the  data samples.

\begin{figure*}[htbp]
  \centering
  \begin{subfigure}[b]{0.48\textwidth}
    \begin{minipage}[b]{\linewidth}
      \centering
      \includegraphics[width=\linewidth, trim={5.35in 0.0in 2.0in  0.0in},clip]{denoising_proc_test_TxConstell-25_16QAM_vBig.png}
      \includegraphics[width=\linewidth, trim={1.5in 0.0in 1.0in  0.0in},clip]{DDPM_test_TxConstell-25_16QAM_vBig.png} 
       \includegraphics[width=\linewidth, trim={1.5in 0.0in 1.0in  0.0in},clip]{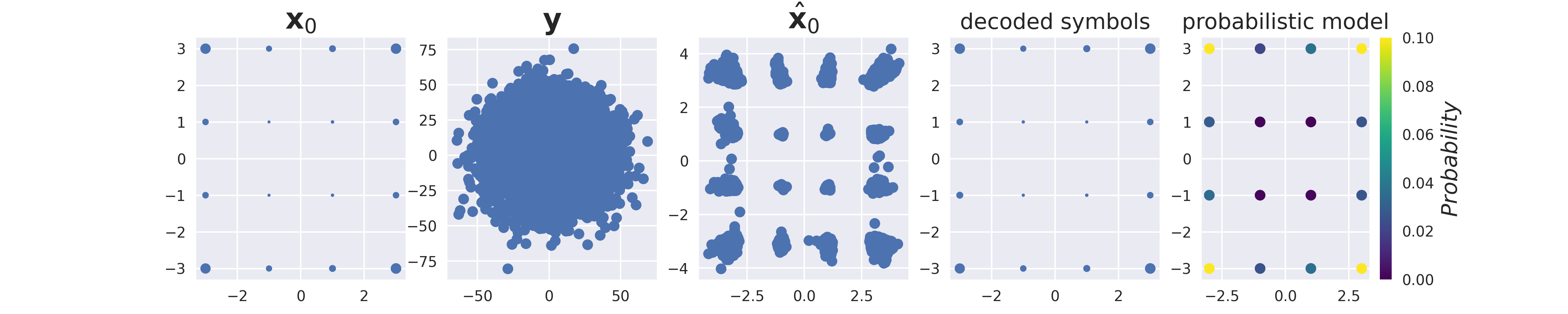}
      \caption{$-25$ dB SNR \vspace{2mm}}
    \end{minipage}
  \end{subfigure}
  \hfill
  \begin{subfigure}[b]{0.48\textwidth}
    \begin{minipage}[b]{\linewidth}
      \centering
      \includegraphics[width=\linewidth, trim={5.35in 0.0in 2.0in  0.0in},clip]{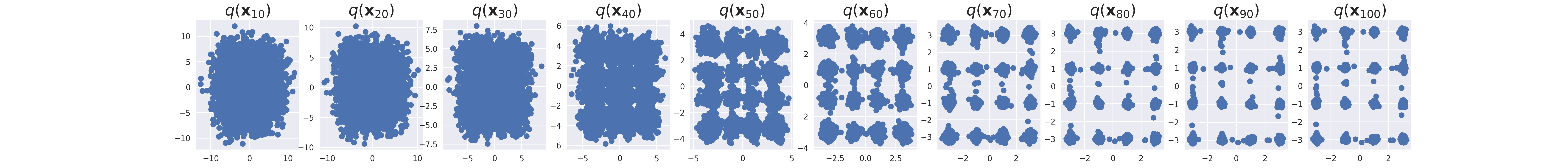}
      \includegraphics[width=\linewidth, trim={1.5in 0.0in 1.0in  0.0in},clip]{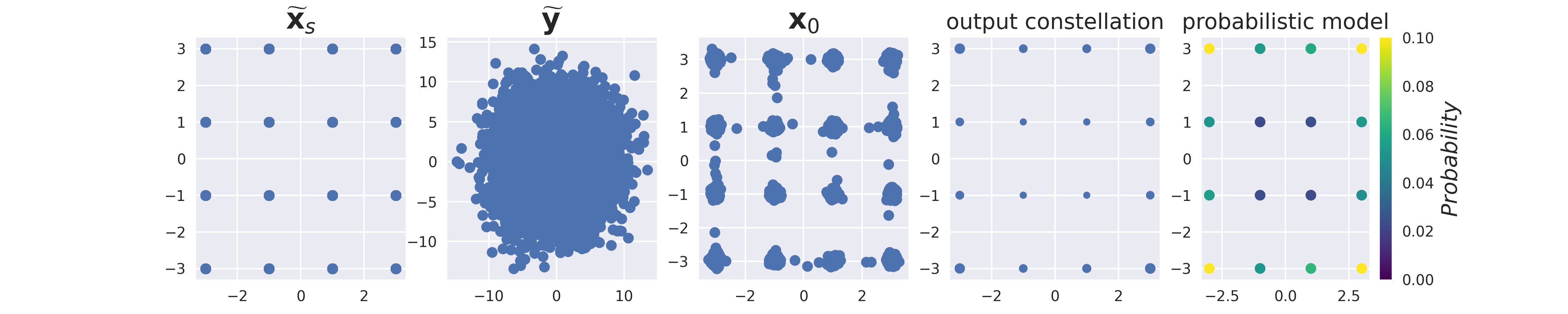} 
      \includegraphics[width=\linewidth, trim={1.5in 0.0in 1.0in  0.0in},clip]{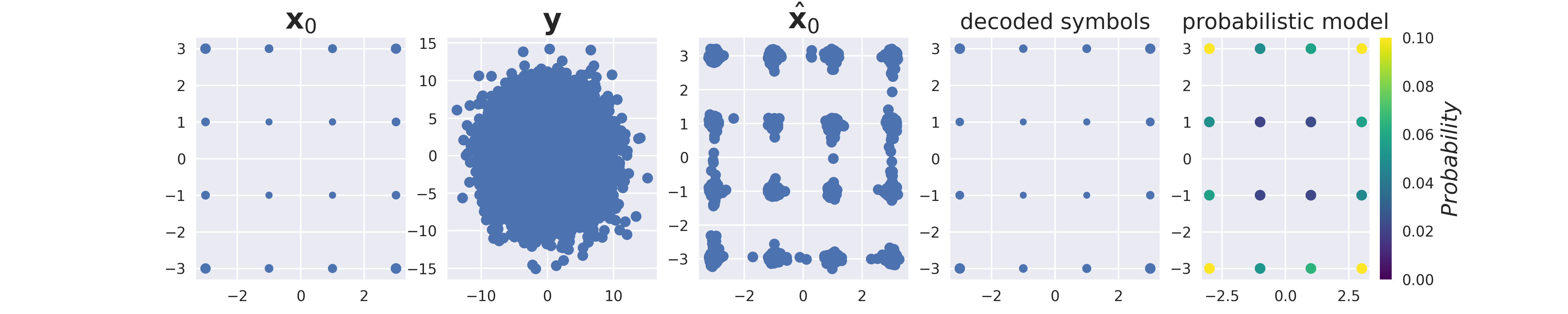} 
      \caption{$-10$ dB SNR \vspace{2mm}}
    \end{minipage}
  \end{subfigure} 
  \begin{subfigure}[b]{0.48\textwidth}
    \begin{minipage}[b]{\linewidth}
      \centering 
      \includegraphics[width=\linewidth, trim={5.35in 0.0in 2.0in  0.0in},clip]{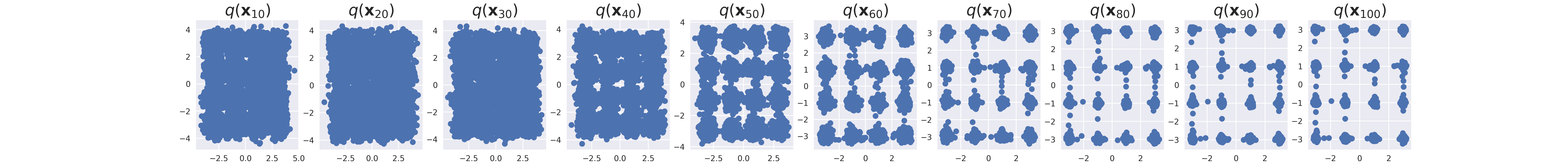}
      \includegraphics[width=\linewidth, trim={1.5in 0.0in 1.0in  0.0in},clip]{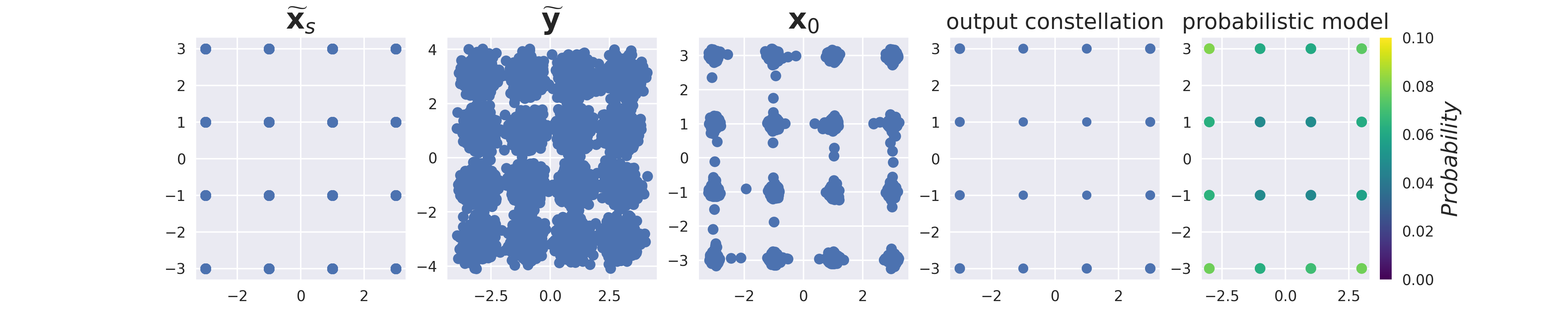} 
     \includegraphics[width=\linewidth, trim={1.5in 0.0in 1.0in  0.0in},clip]{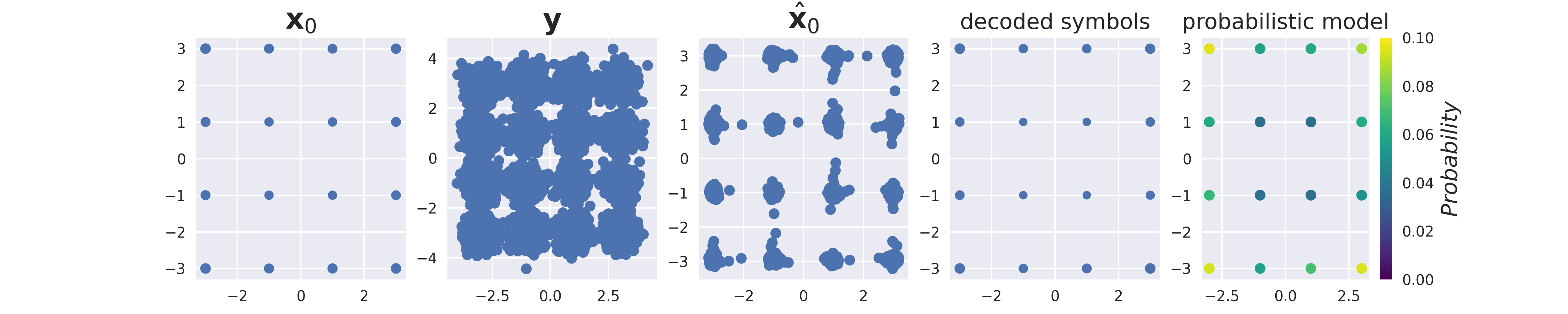} 
      \caption{$10$ dB SNR}
    \end{minipage} 
  \end{subfigure}
  \hfill
  \begin{subfigure}[b]{0.48\textwidth}
    \begin{minipage}[b]{\linewidth}
      \centering
      \includegraphics[width=\linewidth, trim={5.35in 0.0in 2.0in  0.0in},clip]{denoising_proc_test_TxConstell030_16QAM_vBig.png}
      \includegraphics[width=\linewidth, trim={1.5in 0.0in 1.0in  0.0in},clip]{DDPM_test_TxConstell030_16QAM_vBig.png} 
      \includegraphics[width=\linewidth, trim={1.5in 0.0in 1.0in  0.0in},clip]{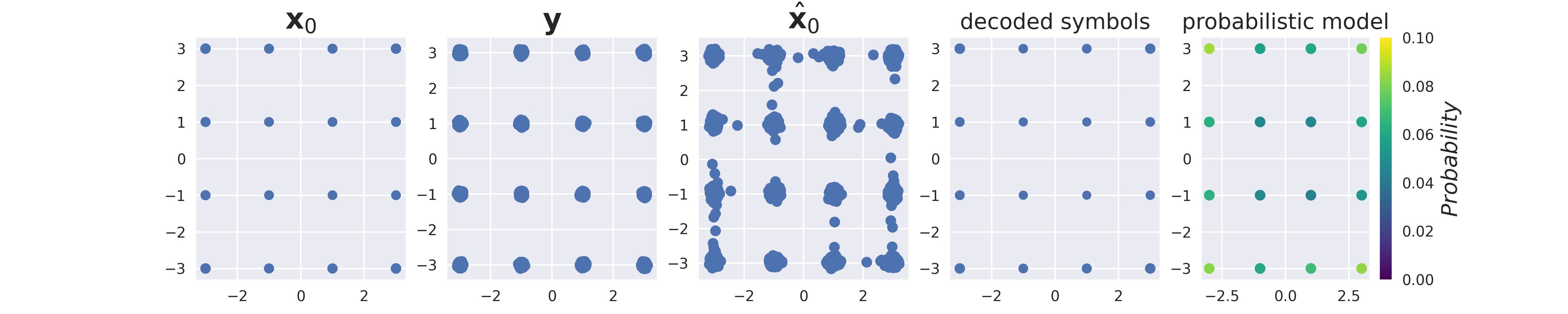} 
      \caption{$30$ dB SNR}
    \end{minipage}
  \end{subfigure} 
\caption{Probabilistic constellation shaping  process at the transmitter, and the symbol  reconstruction at the receiver. 
    From top to bottom, the first row  visualizes  the reverse diffusion process 
    during sampling phase,  
    the second row corresponds to the  probabilistic constellation shaping algorithm,  and the third row corresponds to the reconstruction at the receiver.}\label{fig:denoise}
\end{figure*}

In Fig. \ref{fig:denoise}, we carry out  data visualization for sampling phase  over different SNR values. 
Data visualizations demonstrated in this figure correspond to Algorithms \ref{alg:sampling_Tx} and \ref{alg:sampling_Rx}. 
The first row  corresponds to  data visualization of the reverse diffusion process for generating constellation points, the second row corresponds to the probabilistic  constellation shaping  steps that are performed  at the transmitter (Algorithm \ref{alg:sampling_Tx}), and the third row corresponds to the reconstruction at the receiver (Algorithm \ref{alg:sampling_Rx}).   This is repeated for different SNRs.        
According to data visualizations for  the reverse diffusion process,  
our results show  that even when the power of noise is more than $300$ times of the signal ($\Gamma = -25$) we can have a clear understanding of the underlying  constellation scheme at the end of running the diffusion model. Also, having higher SNRs results in more clear generation and reconstruction of constellation samples.  
To explain    
the second and the third rows of Fig. \ref{fig:denoise} for each SNR value,  we can provide the following discussions.   
{For each SNR level, we  start with $N_s = 10000$ uniformly-sampled  constellation points (the first sub-figure), synthetically add  Gaussian  noise  to  the samples, and run the DDPM 
according to Algorithm \ref{alg:sampling_Tx}. We then map the generated  samples  (the third sub-figure) to the nearest constellation point in set $\mathcal{X}_c$ which results in the forth sub-figure, in which the size of the points is proportional to their probabilities of occurrence.  Finally, by counting the number of occurrence (generation) of each constellation  symbol  generated  at the output of the DDPM block, the probabilistic model of the  constellation points is obtained (the fifth sub-figure).  
Afterward,  
information symbols can be mapped onto the constellation points based on the distribution obtained from the probabilistic model. The probabilistically-shaped information symbols are passed through  the communication channel and received by the receiver end as denoted by $\boldsymbol{y}$.    
The receiver then runs the DDPM model based on Algorithm \ref{alg:sampling_Rx}, reconstructs, and decodes the constellation symbols, as shown in the third row of Fig. \ref{fig:denoise} for each SNR level.   
Comparing the output of our probabilistic constellation generation algorithm (the second row) and the reconstructed symbols at the receiver (the third row), we can observe that the idea of  
mimicking the functionality of receiver for shaping  the constellation symbols (addressed in Section \ref{subsec:proposed_approach}) has helped the transmitter generate symbols that are quite similar to the ones that are actually reconstructed by the receiver. This improves the communication performance by significantly  decreasing the mismatch between the way the transmitter conveys   the information, and the way the receiver  decodes the  symbols.  
This ``similarity'' is quantitatively measured  in terms of mutual information and cosine similarity  in the subsequent  figures.}

{ 
To further study  the behaviour of our proposed  scheme over different SNR values, we can provide the following  explanations.  
According to the results of Fig. \ref{fig:denoise},    when the communication system is experiencing low-SNR regimes (e.g., $-25$ dB or $-10$ dB in the figure),   the probabilistic model at the output of transmitter's DDPM demonstrates a non-uniform distribution over constellation points, with higher probabilities assigned to the  points that are at the furthest distance from each other in the constellation geometry. This is aligned with what we intuitively expect  from a communication system under low-SNR regimes  to   frequently  map information bits to constellation symbols that are far apart from each other, in order to decrease the decoding error.     
Increasing the SNR, we can see from the figure that the probabilistic shaping tends to uniform distribution, which is also  aligned with one's intuition about communication systems.}  
We finally mention that our results for 64-QAM scenario also demonstrated  quite the same results  
where we do not repeatedly mention them here for the sake of brevity.

\subsection{Performance Evaluation and Resilience }\label{ssec:evaluations_performance}
In this subsection, we present  quantitative evaluations of our proposed scheme  and show its \emph{resilience} against  low-SNR regimes, as well as \emph{out-of-distribution (OOD)
robustness}  for  non-Gaussian  noise.  We also compare our results with different benchmarks.

\begin{figure} 
\centering
\includegraphics
[width=3.5in,height=2.4in,trim={0.2in 0.0in 0.45in  0.5in},clip]{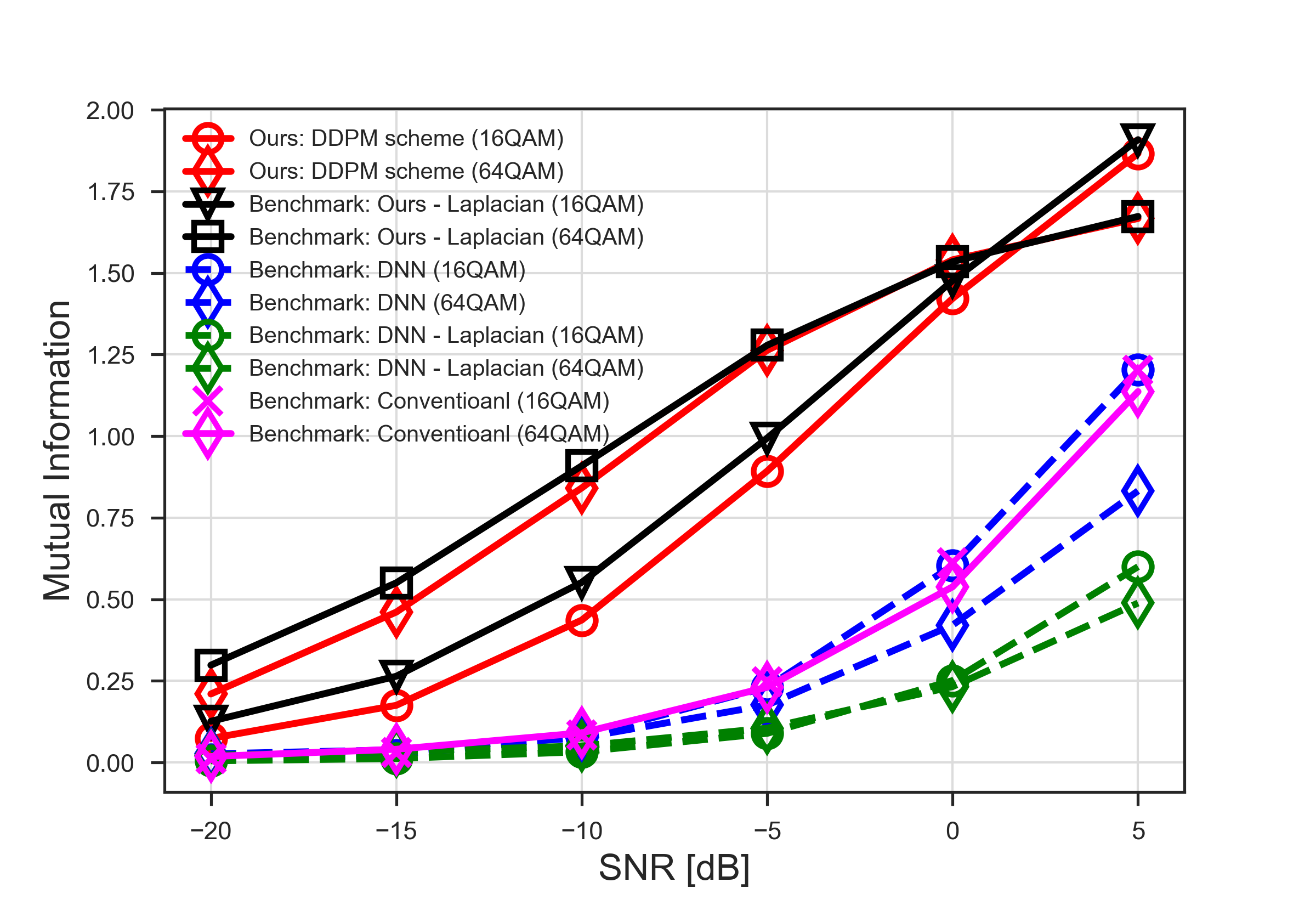}
\vspace{-3mm}
\caption{Mutual information  between the   
generated symbols at transmitter
and  the reconstructed ones at the receiver for AWGN channel and non-Gaussian noise.   
}
	\label{fig:MI}
 \vspace{0mm}
\end{figure}

Fig. \ref{fig:MI} demonstrates the mutual information  between the   generated symbols at the output of  transmitter's DDPM  (i.e., the channel input),  and  the reconstructed ones at the receiver. 
The mutual information metric can be considered as a fundamental measure to quantitatively  assess  the ``mutual similarity'' between the  communication nodes.   
For this experiment, we consider both cases of additive white Gaussian noise (AWGN) channel and also  non-Gaussian noise to highlight the OOD performance of our scheme.   
For the DNN benchmark, we consider a learning-based  scenario  with trainable constellation layer and neural demapper \cite{constell}.\footnote{Generally speaking, this benchmark  is already  implemented  by Sionna (an open source library developed and used by NVIDIA to carry out 6G research), which is referred to trainable constellation,  
and it is considered as one of the key advantages of Sionna  for ML-based research \cite{Sionna}.}   
The DNN benchmark has three linear layers with $64$ neurons and rectified linear unit (ReLU) activation functions \cite{Sionna}. 
We note that based on our ablation studies, increasing the depth of the  DNN, or the number of hidden neurons does not result in a significant  improvement in its  performance. 
For both $16$- and $64$-QAM benchmarks, we considered $5000$ training iterations  
with Adam optimizer \cite{Sionna}.   
Considering the general behaviour of mutual information versus SNR in this figure, we can observe  that by increasing the SNR level, it becomes more straightforward for both the transmitter and the receiver to denoise and generate the symbols, which is also aligned with one's intuition.  Hence, the mutual information increases with the SNR as can be observed from the figure.  The increasing  trend of mutual information vs. SNR also   applies to the DNN benchmark, since a typical  DNN can also perform the inference with higher accuracy  when given less noisy input.    

Fig. \ref{fig:MI}  highlights the performance of our proposed  DDPM-based  scheme compared to the DNN-based approach and the conventional  uniform shaping.    While the conventional and DNN benchmarks  
do not exhibit  any significant   performance in SNR ranges below $-5$ dB,   
our scheme achieves a  mutual information of around $1$ bit for 16-QAM case, and $1.25$ bits for 64-QAM case,  respectively.  
In addition, one can observe  a threefold improvement  in terms of mutual information between the communication sides compared to DNN-based benchmark at $0$ dB SNR.  
These results clearly highlight that our main goal in realizing the ``mutual understanding'' among communication parties has  been successfully achieved, and thanks to this understanding,  the system is resilient under low-SNRs.     
To show the robustness of our scheme for OOD performance, we study the scenario of communication  channels with non-Gaussian noise \cite{twelve_6G}.     
Specifically, for  this experiment  we consider additive Laplacian noise with the same variance as that of AWGN scenario.   Remarkably, although we do not re-train our diffusion model with  Laplacian noise,  the performance of  our DDPM-based approach does not change  
under this OOD scenario, and the resultant  mutual information curves follow the case of in-distribution scenario.  However, we can see from the figure that the DNN benchmark experiences performance degradation under non-Gaussian assumption,  although  we also re-trained it with Laplacian noise.    
For example, for 16-QAM geometry and $5$ dB SNR, the mutual  information of DNN benchmark decreases by more than $50\%$ in Laplacian case.

\begin{figure} 
\centering
\includegraphics
[width=3.5in,height=2.45in,trim={0.3in 0.0in 0.6in  0.6in},clip]{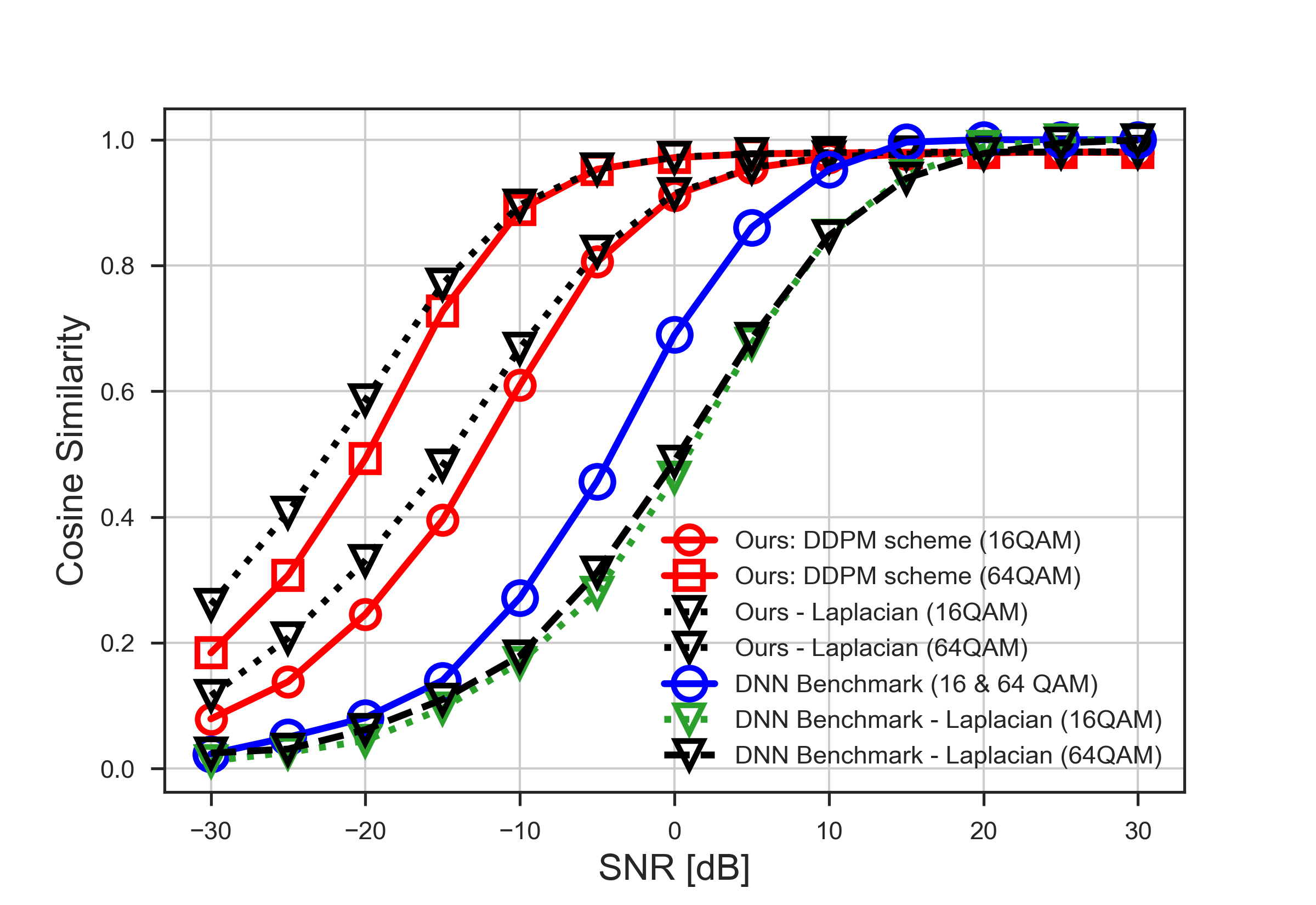}
\vspace{-4mm}
\caption{Cosine similarity  between transmitter's generated  symbols 
 and  the reconstructed ones at the receiver.  
}
	\label{fig:CS}
 \vspace{0mm}
\end{figure}

In Fig. \ref{fig:CS}, we further study  the  performance  of  our proposed scheme in terms of cosine similarity, over a wide range of SNR values from  $-30$ dB to $30$ dB.  
The cosine similarity  is a widely-adopted measure in data science to quantify the similarity between two vectors, $\mathbf{x}_0$ and  $\widehat{\mathbf{x}}_0$ \cite{3gpp_CosineSim}.  It can be formulated as 
\begin{align}
\mathsf{CSIM} = \frac{\mathbf{x}_0^{\mathsf{T}} \widehat{\mathbf{x}}_0}{||  \mathbf{x}_0||  
\hspace{1.5mm} 
|| \widehat{\mathbf{x}}_0||}.   
\end{align}
Here, we calculate the cosine similarity between the output constellation symbols at the transmitter, and the reconstructed  ones at the receiver.   
Since  we have 2-D constellation points as I/Q samples, we take the cosine similarity of I and Q samples separately, and then 
normalized it to $1$.    
Notably, the figure  highlights that our proposed approach  leads to competitive  in- and out-of-distribution performance when compared to  DNN-based   benchmarks.   
For example,  for $-5$ dB SNR and 64-QAM geometry, the system 
maintains  the ``mutual  similarity''  between the communication parties such that  more than $50\%$ improvement could be observed  in terms of cosine similarity  compared to  the DNN benchmark.  
Moreover, while the performance of  DNN benchmark degrades when having non-Gaussian noise,  our scheme achieves the same (or even better) results and maintains its robustness.

\begin{figure*}[t!]
    \centering
    \begin{subfigure}[t]{0.5\textwidth}
        \centering
\includegraphics[width=3.4in,height=2.2in,trim={0.3in 0.0in 0.0in  0.0in},clip]
    {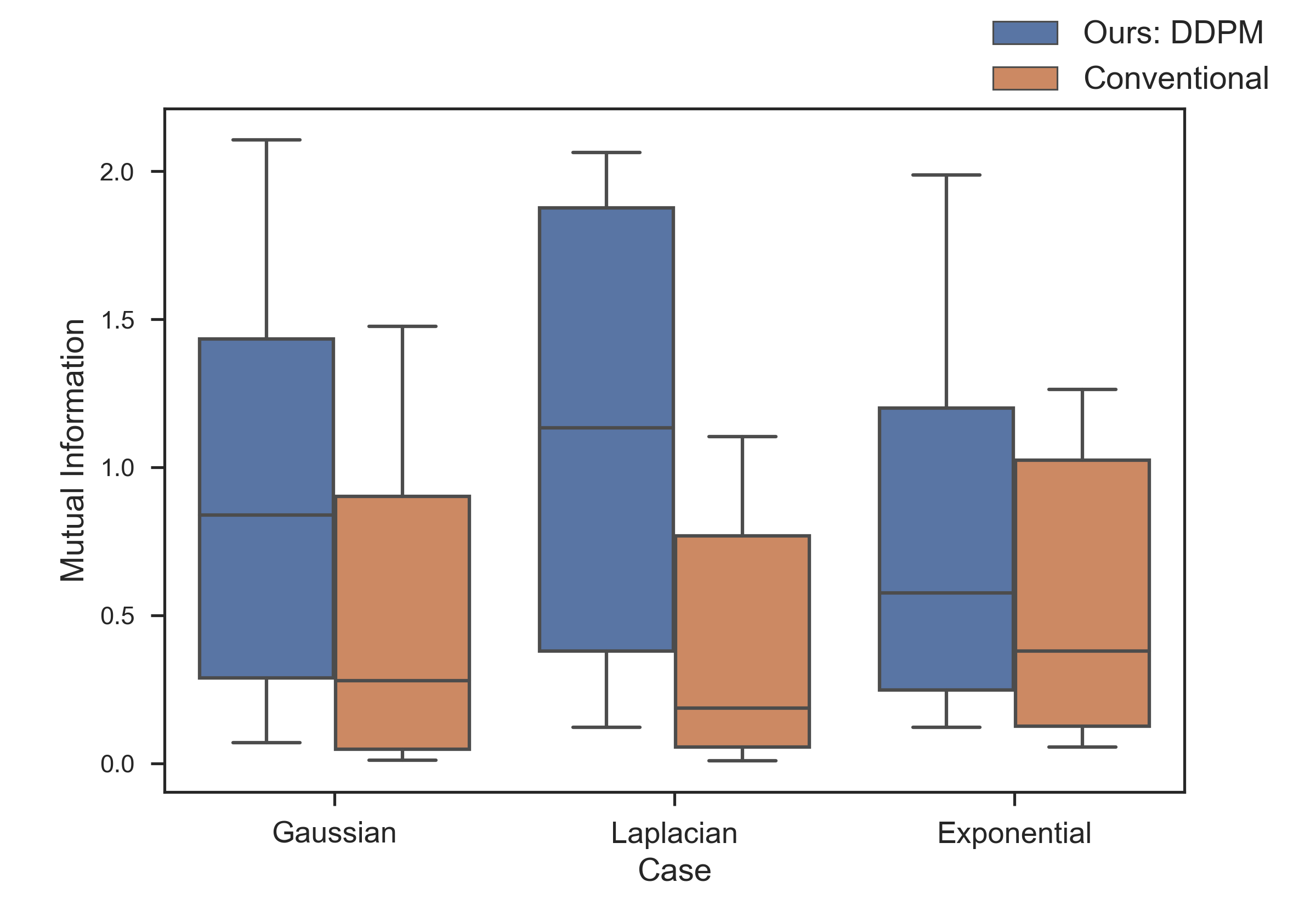}
        \caption{$16$-QAM}
    \end{subfigure}%
    \begin{subfigure}[t]{0.5\textwidth}
        \centering
\includegraphics[width=3.4in,height=2.2in,trim={0.3in 0.0in 0.0in  0.0in},clip]
        {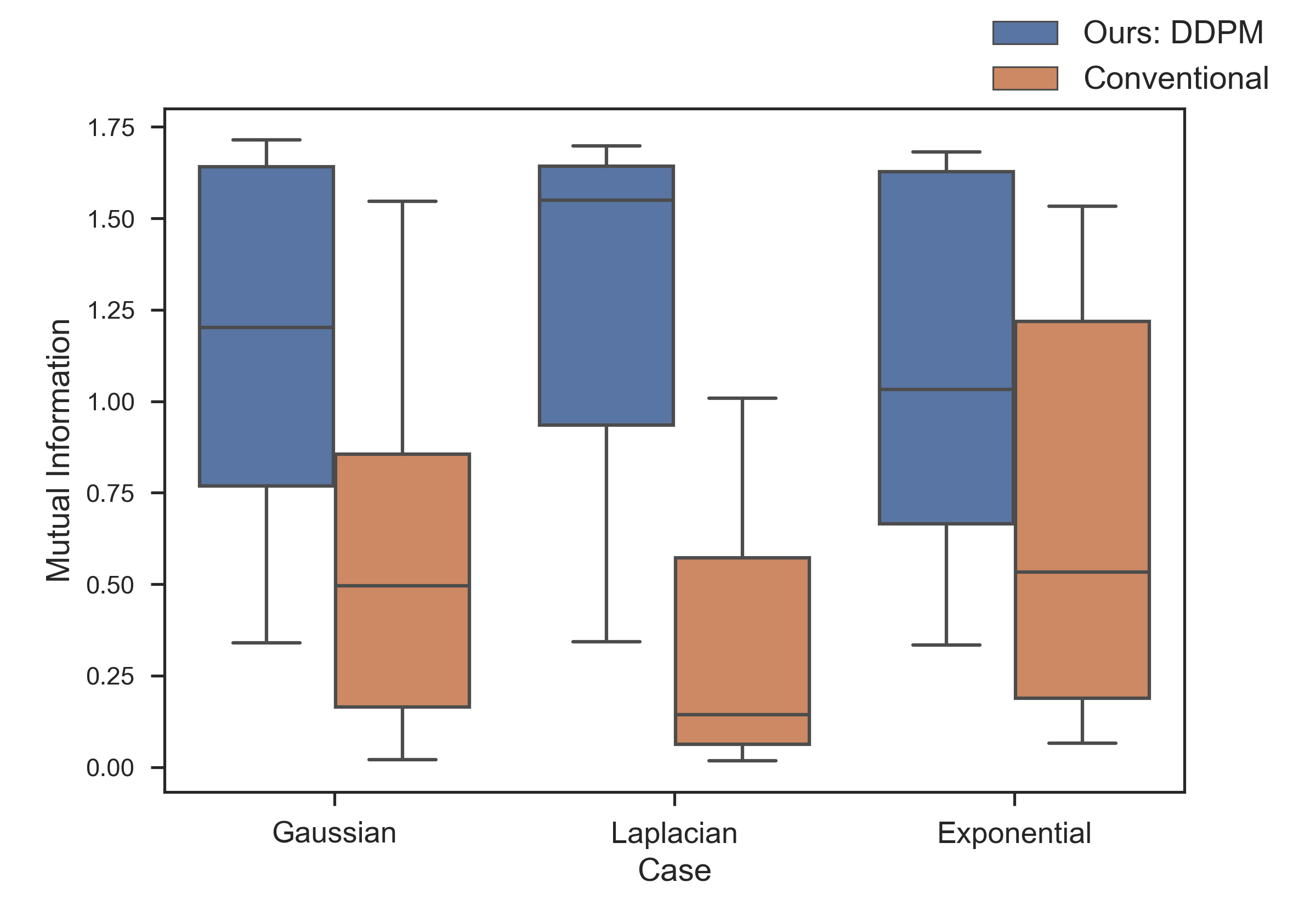}
        \caption{$64$-QAM}
    \end{subfigure}
    \caption{ Performance of our proposed scheme in terms of mutual information for both in- and out-of-distribution scenarios.}
    \label{fig:boxplot_MI}
\end{figure*}

\begin{figure*}[t!]
    \centering
    \begin{subfigure}[t]{0.5\textwidth}
        \centering
\includegraphics[width=3.4in,height=2.2in,trim={0.3in 0.0in 0in  0.0in},clip]
    {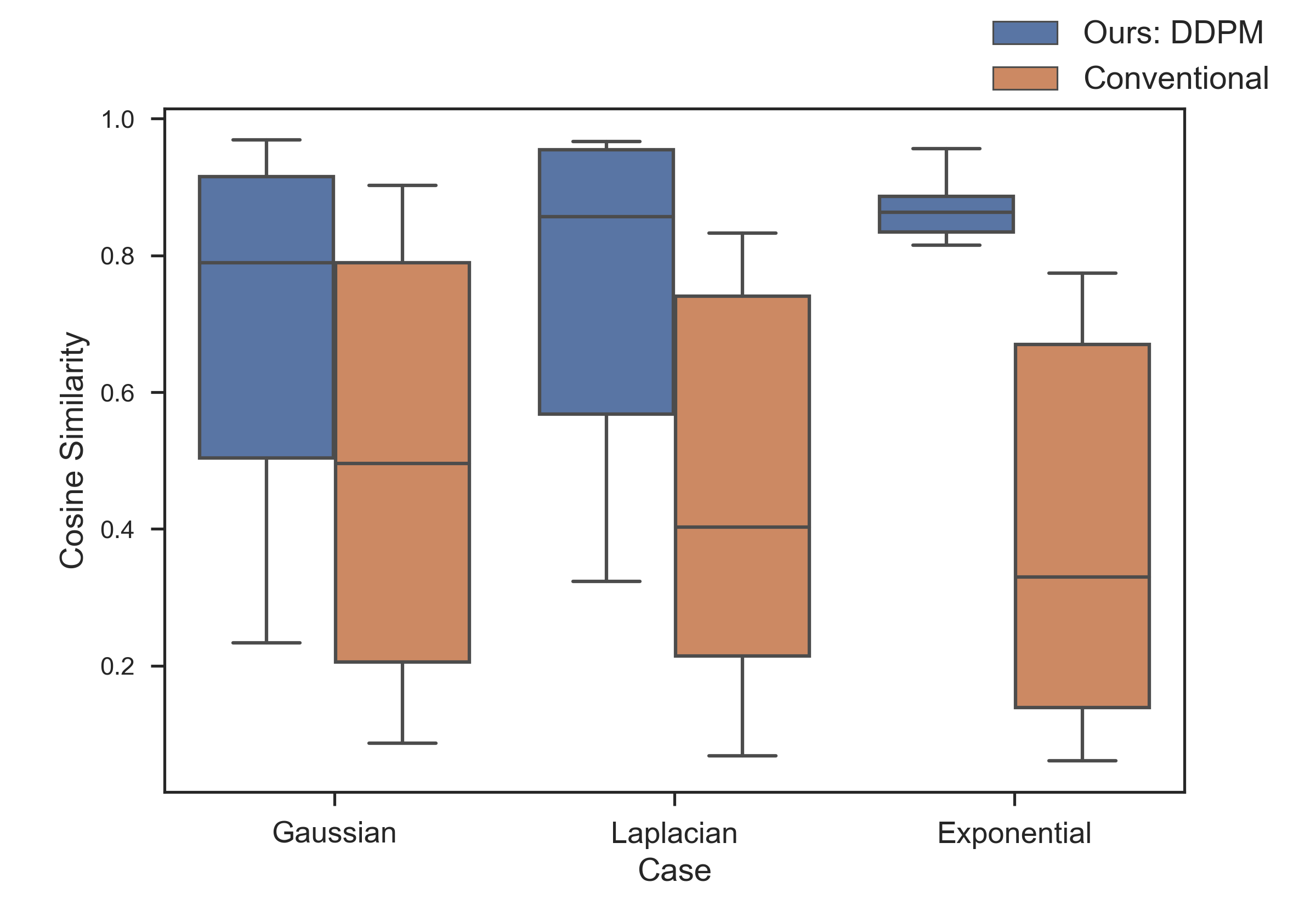}
        \caption{$16$-QAM}
    \end{subfigure}%
    \begin{subfigure}[t]{0.5\textwidth}
        \centering
\includegraphics[width=3.4in,height=2.2in,trim={0.3in 0.0in 0.0in  0.0in},clip]
        {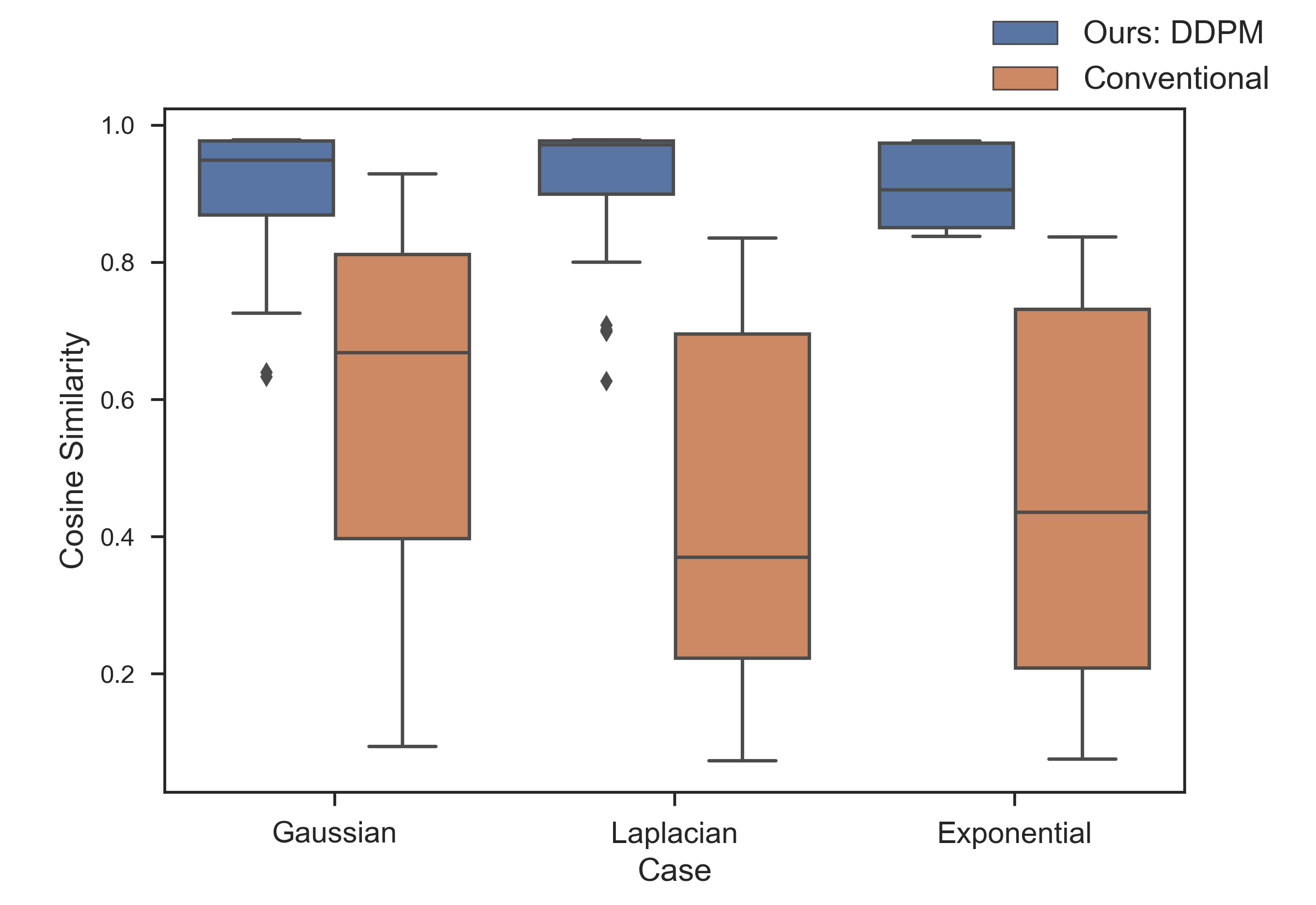}
        \caption{$64$-QAM}
    \end{subfigure}
    \caption{Performance of our proposed scheme in terms of cosine similarity  for both in- and out-of-distribution scenarios.}
    \label{fig:boxplot_CS}
\end{figure*}

We now consider a different setup, in order to take into account  the variations of wireless channel (equivalently the channel SNR in our system model),  over the transmission  time-slots.   This is different from the previous simulation setups, for  which we set a fixed SNR value, simulated the communication system, and derived the performance metrics for each SNR.    
To this end, we run the probabilistic shaping algorithm multiple times,  each  time considering a different channel SNR  which is   chosen randomly from the set of SNR levels $\{-20, -19, \cdots, 9, 10\}$. Now we are interested in the distribution of our  performance metrics derived  over those  realizations.

Figs. \ref{fig:boxplot_MI} and \ref{fig:boxplot_CS}  demonstrate, respectively,  the distribution of mutual information, and the cosine similarity achieved  over $30$ realizations of our diffusion-based  probabilistic constellation shaping (each with empirical sample size of $N_s = 10000$), versus the conventional (non-learning) uniform shaping. The distributions are illustrated as  box-plots so that we can  graphically show the locality and skewness, as well as the variation in the performance metrics derived over time. 
The results are obtained for three different scenarios, i.e., the in-distributions scenario with Gaussian noise, and the OOD scenario with Laplacian and exponential noise \cite{twelve_6G}, without re-training the model.  
We can infer from the figures that our proposed scheme significantly outperforms the conventional uniform shaping in different scenarios, as for most of the plots, the median line of the performance metrics for our scheme lies outside the box of the corresponding  conventional baseline.  
Furthermore, our proposed scheme consistently outperforms the conventional scheme, achieving higher median and maximum performance metrics in all of the scenarios.  
We can also see from the figures that  the generative AI-based approach even performs better with the increase in the modulation order (which is also aligned with the per-SNR results in Figs. \ref{fig:MI} and \ref{fig:MI}), while the conventional approach experiences performance degradation in $64$-QAM case.  
Considering the  OOD scenario with Laplacian noise, the figures imply that the performance of the conventional baseline decreases over time, in both $16$-QAM  and $64$-QAM cases. This is due to the fact that the conventional  solutions  are optimized with the underlying assumption of Gaussian distribution, and  they typically  lead to non-optimal solutions for other types of noise \cite{twelve_6G}.       
On the other hand, our DDPM-based approach  does not show any performance degradation in terms of mutual information, nor in terms of cosine similarity.  
We finally mention that for the special case of  exponential noise, 
the conventional baseline might perform better  than the Laplacian scenario. This is due to the fact that with exponential noise, the system observes a one-sided distribution for the additive noise, and hence, it would be  easier to decode the symbols than the case of Laplacian noise.

\section{Conclusions}\label{sec:concl} 
In this paper,  we have proposed DDPM-based  probabilistic constellation shaping for wireless communications.   
exploiting  the ``denoise-and-generate'' characteristic  of diffusion models, we have offered a radically different approach for PHY design  based on generative AI.    
The key idea  was that   the transmitter mimics the way the receiver would do to reconstruct (regenerate) the symbols out of noisy signals, realizing   
``reciprocal understanding'' to reduce  the mismatches  among communication parties.   
Our results 
have highlighted the performance of our scheme  compared to  other bechmarks, while providing \emph{network resilience} under  low-SNR regimes and  non-Gaussian noise.  We have achieved  $30\%$ improvement in terms of cosine similarity and a threefold improvement in terms of mutual information compared to  DNN-based approach for 64-QAM geometry. 
We believe that our results in this paper can pave the way towards a new paradigm of  generative AI-based signal design for the future generations  of communication systems.


\end{document}